\documentclass[prd,preprintnumbers,nofootinbib,superscriptaddress,twocolumn]{revtex4-1}
\usepackage{graphicx}
\usepackage{amsmath}
\usepackage{amssymb}
\usepackage{mathtools}
\usepackage[utf8]{inputenc}
\usepackage[bookmarksopen,bookmarksnumbered]{hyperref}
\usepackage{color}
\usepackage[usenames,dvipsnames]{xcolor}
\usepackage[normalem]{ulem}
\usepackage{soul}
\usepackage{units}
\usepackage{rotating}

\allowdisplaybreaks

\def\xleft{\mathopen{}\left}

\DeclareRobustCommand*\diff[2][]{%
   \mathop{
     \mathrm{d}^{#1}
     \mskip-0.2\thinmuskip
    #2}\nolimits
}

\newcommand{\T}[1]{\boldsymbol{#1}_{\text{T}}}
\newcommand{\Tsc}[1]{#1_{\text{T}}}

\newcommand\BIBvol[1]{\textbf{#1}}

\definecolor{dpmagenta}{rgb}{0.8, 0.0, 0.8}

\newcommand{\no}{\nonumber \\}
\newcommand{\bmax}{b_{\rm max}}
\newcommand{\bmin}{b_{\rm min}}

\newcommand\bstarsc{b_*}

\newcommand\bone{b_c}
\newcommand\mubstar{\mu_{\bstarsc}}
\newcommand\muQ{\mu_Q}

\newcommand{\appor}[1]{{\rm T}_{\rm #1}}

\newcommand\mub{\mu_b}

\newcommand{\cs}[2]{ \Gamma^{#2}(\Tsc{q},Q_{#1}) }

\newcommand{\as}[2]{{\rm AY}^{#2}(\Tsc{q},Q_{#1})}
\newcommand{\asnew}[2]{{\rm AY}_{\rm New}^{#2}(\Tsc{q},Q_{#1};\eta,C_5)}
\newcommand{\fixo}[2]{{\rm FO}^{#2}(\Tsc{q},Q_{#1})}
\newcommand{\TT}[2]{W^{#2}(\Tsc{q},Q_{#1})}
\newcommand{\TTnew}[2]{W_{\rm New}^{#2}(\Tsc{q},Q_{#1};\eta,C_5)}
\newcommand{\TTa}[2]{W_{\rm a}^{#2}(\Tsc{q},Q_{#1};\eta,C_5)}

\newcommand{\YY}[2]{Y^{#2}(\Tsc{q},Q_{#1})}
\newcommand{\YYnew}[2]{Y_{\rm New}^{#2}(\Tsc{q},Q_{#1};\eta,C_5)}
\newcommand{\mucirc}{\bar{\mu}}
\newcommand{\mucircm}{\mu_c}

\newcommand{\arrowcom}[1]{\textcolor{red}{\textbf{$\Longrightarrow$ #1}} \\}

\def\??#1{\mbox{}\\\arrowcom{#1}}
\long\def\!!#1{{\color{red} #1}}

\begin{document}


\title{Relating Transverse Momentum Dependent and Collinear Factorization Theorems in a Generalized Formalism}
  
\preprint{JLAB-THY-16-2245}  
\author{J. Collins}
\email{jcc8@psu.edu}
\affiliation{%
  Department of Physics, Penn State University, University Park PA 16802, USA
}
\author{L. Gamberg}
\email{lpg10@psu.edu}
\affiliation{Science Division, Penn State University Berks, Reading, Pennsylvania 19610, USA}
\author{A.~Prokudin}
\email{prokudin@jlab.org}
\affiliation{Science Division, Penn State University Berks, Reading, Pennsylvania 19610, USA}
\affiliation{Theory Center, Jefferson Lab, 12000 Jefferson Avenue, Newport News, VA 23606, USA}
\author{T.~C.~Rogers}
\email{tedconantrogers@gmail.com}
\affiliation{Department of Physics, Old Dominion University, Norfolk, VA 23529, USA}
\affiliation{Theory Center, Jefferson Lab, 12000 Jefferson Avenue, Newport News, VA 23606, USA}
\author{N.~Sato}
\email{nsato@jlab.org}
\affiliation{Theory Center, Jefferson Lab, 12000 Jefferson Avenue, Newport News, VA 23606, USA}
\author{B.~Wang}
\email{bowenw@mail.smu.edu}
\affiliation{Department of Physics, Old Dominion University, Norfolk, VA 23529, USA}
\affiliation{Theory Center, Jefferson Lab, 12000 Jefferson Avenue, Newport News, VA 23606, USA}

\date{August 15, 2016}

\begin{abstract}
    We construct an improved implementation for combining 
    transverse-momentum-dependent (TMD) factorization
    and collinear factorization.  TMD factorization is suitable for
    low transverse momentum physics, while collinear factorization is
    suitable for high transverse momenta and for a cross section
    integrated over transverse momentum.  The result is a modified
    version of the standard $W+Y$ prescription traditionally used in
    the Collins-Soper-Sterman (CSS) formalism and related approaches.
    We further argue that questions regarding the shape and
    $Q$-dependence of the cross sections at lower $Q$ are largely
    governed by the matching to the $Y$-term.  
\end{abstract}

\maketitle

\section{Introduction}
\label{sec:intro}
Much of the literature on the TMD-factorization formalism  is based on methods like those of Collins Soper and Sterman (CSS)~\cite{Collins:1981uk,Collins:1981uw, Collins:1984kg, Collins:2011qcdbook} 
 where, traditionally,
applications have been at very high scales. 
The formalism involves a factorization with
TMD parton densities and/or
fragmentation functions together with evolution equations and associated properties like universality.  
TMD correlation functions have attracted interest, both for their 
usefulness in perturbative calculations, 
and for their potential to yield information about underlying 
non-perturbative QCD structures.
Results with essentially the same or a related
structure are also found in SCET
~\cite{Becher:2010tm,Echevarria:2012pw,Rothstein:2016bsq}.
In this paper, we focus on the CSS formalism 
 and  its updated version in~Ref.~\cite{Collins:2011qcdbook}. 

TMD correlation functions are most useful for $\Tsc{q} \ll Q$, where $\Tsc{q}$ is the relevant 
transverse momentum and $Q$ is the overall hard scale. 
When $\Tsc{q}$ is of order $Q$, the cross section does not factor into TMD correlation functions, but normal collinear
factorization applies.  It is, of course, necessary to be able to
analyze cross sections over the whole range of $\Tsc{q}$ including
intermediate transverse momenta.  To this end, CSS organized the cross
section into an additive form, $W+Y$, where $W$ is the pure TMD factorization
term and $Y$ is a correction term using collinear factorization.  $W$ dominates in the limit 
of small $\Tsc{q}/Q$ while $Y$ is a correction for large $\Tsc{q}/Q$. This
was designed with the aim to have a formalism that is valid to leading
power in $m/Q$ uniformly in $\Tsc{q}$; here $m$ is a typical hadronic
mass scale.

However, it has become increasingly clear that the original
CSS $W+Y$ method is not sufficient for modern TMD applications. One reason is that there is a growing number of 
lower-$Q$ phenomenological studies focused on the intrinsic 
transverse motion related to nonperturbative binding and nucleon structure.
The advantages of the usual $W+Y$ 
decomposition are clearest when $Q$ is large enough
that there is a broad intermediate range of transverse momentum
characterized by $m \ll \Tsc{q} \ll Q$; that is, there is a range where $\Tsc{q}/Q$ is
sufficiently small that TMD factorization is valid to good accuracy,
while $m/\Tsc{q}$ is also sufficiently small that collinear
factorization is simultaneously valid. However, at lower phenomenologically interesting values of
$Q$, neither of these ratios is necessarily very small. 
Some other difficulties will be summarized below.  These
  particularly concern the ability of the original $W+Y$ method to
  properly match collinear factorization for the cross section
  integrated over $\T{q}$.

The problems create practical difficulties for studies specifically devoted to extracting 
and analyzing non-perturbative transverse momentum dependence. For such applications, the relevant experiments 
often involve hard scales of only a few GeV. The phase space of $\Tsc{q}$ has a narrow transition window between 
a solidly perturbative transverse momentum region (where $\Tsc{q}\simeq O(Q)$) and a 
non-perturbative region (where $\Tsc{q} \simeq O(m)$), making the treatment of matching the perturbative and nonpertubative content in the intermediate region rather delicate. 
A classic analysis of the issues concerning the  matching of the TMD factorization
and collinear factorization was given  by Arnold and Kauffman
\cite{Arnold:1990yk}, and  more recently  in Refs.~\cite{Guzzi:2013aja,Su:2014wpa,Boglione:2014oea,Boer:2015uqa}. 
See especially Sec.~2.6 of Ref.~\cite{Guzzi:2013aja} for a recent overview of many of the issues to be discussed in this paper.

Over the past several years, most theoretical attention in TMD physics has been focused on 
the details of evolution of the $W$-term and its associated TMD correlation functions. 
However, particularly with recent results like~\cite{Su:2014wpa,Boglione:2014oea,Boer:2015uqa}, 
it is  evident that a satisfactory treatment of non-zero $\Tsc{q}/Q$ corrections 
and the matching to $\Tsc{q} \gtrsim Q$ is important since it relates various phenomenological analyses to TMD theory.  This is especially the case in efforts to 
interpret transverse momentum spectra in terms of hadronic structure, where a detailed separation and 
identification of large and small $\Tsc{q}/Q$ behavior and its potential interplay is important. 

Generally, to get results that are valid over all $\Tsc{q}$ we need to
combine the information given by TMD factorization and by collinear
factorization.  TMD factorization is appropriate for $\Tsc{q} \ll Q$;
its accuracy degrades as $\Tsc{q}$ increases and eventually it does
not give even a qualitatively correct account of the cross section.
Collinear factorization is valid in two ways.  One is for the cross
section differential in $\Tsc{q}$ with $\Tsc{q} \sim Q$; the accuracy
degrades as $\Tsc{q}$ decreases, and collinear factorization becomes
entirely inapplicable for the differential cross section once
$\Tsc{q}$ is of order $m$ or smaller.  But collinear factorization is
also valid for the cross section integrated over $\T{q}$.  

In this article, we argue for an enhanced formalism. As already
stated, the $W+Y$ formalism as given by CSS was designed to combine
the best of TMD and collinear factorization at intermediate $\Tsc{q}$.
What was not done was to adjust the formalism to work nicely also for
the cross section integrated over all $\T{q}$.  We summarize an
interconnected set of problems as follows:
\begin{itemize}
\item A standard way of presenting the $W$ term, with the solution to
  the evolution equations, is as a Fourier transform from a transverse
  coordinate $\Tsc{b}$ to transverse momentum.  When $\Tsc{b} \to 0$,
  the $\Tsc{b}$-space integrand $\tilde{W}(\Tsc{b})$ goes to zero.
  (See Appendix~\ref{sec:Wzero}.) Therefore, the integral over all
  transverse momentum of the corresponding momentum-space contribution
  $W(\Tsc{q})$ is zero.  Now, at small $\Tsc{q}$, $W(\Tsc{q})$ is the
  dominant TMD-factorized contribution to the cross section, and is
  necessarily positive. Therefore, at some larger $\Tsc{q}$, the
  $W(\Tsc{q})$ term must become negative. By construction, the $Y$
  term compensates to give the physical positive cross section, so
  this is not a problem in principle. However, if $W$ becomes
  \emph{large} and negative at $\Tsc{q} \sim Q$, the $Y$ term becomes
  large and positive, so the formalism involves implementing a
  cancellation of two large quantities.  This can enormously magnify
  the effects of truncation errors in perturbative quantities, since
  these have different structures in $W$ and $Y$.

\item In pure parton-model treatments of TMD functions, the transverse
  momentum integral of the $W$-term gives the collinear factorization
  parton model for the cross section integrated over $\T{q}$. The
  previous item shows that, at least within the original CSS approach,
  this connection is not merely subject to higher-order perturbative
  corrections, but is totally lost.

\item In real QCD, consider the cross section integrated over all
  $\T{q}$;  it is of the form of factors
  of collinear parton densities and/or fragmentation functions at
  scale $Q$ convoluted with hard scattering that is expanded in powers
  of $\alpha_s(Q)$. The lowest order for the integrated cross section itself is
  correctly given by a perturbative expansion of the hard scattering,
  with the first term being zeroth order in $\alpha_s(Q)$
  (concentrated at $\Tsc{q}=0$). We can try doing this for all 
   quantities in
  \begin{equation}
  \label{eq:int.sigma}
    \int \diff[2]{\T{q}} 
    \frac{ \diff{\sigma} }{ \diff[2]{\T{q}} \dots }
    =
    \int \diff[2]{\T{q}} W
    + \int \diff[2]{\T{q}} Y.
  \end{equation}
   Since the integral over $W$ is zero,
  the integrated cross section is given by the integral over $\T{q}$
   of the $Y$ term.  But the CSS construction 
   of the $Y$ term shows
  that its lowest term is the same order as for collinear
  factorization for the differential cross section, which is first
  order in $\alpha_s(Q)$~\cite{Collins:1984kg} .

  We thus have a paradox: a mismatch of orders in $\alpha_s(Q)$
  between the left and right hand sides of Eq.\ (\ref{eq:int.sigma}).
  The real source of the paradox and an indications of what to do
  about are indicated next.

\item The zero value of $\int \diff[2]{\T{q}} W$ is not obtained from
  a fixed order perturbative application of collinear factorization to
 $\tilde{W}(\Tsc{b},Q)$ at $\Tsc{b}=0$, but from the solution of
  evolution equations for $\tilde{W}$, as seen in Eq.\
  (\ref{eq:finalevolved}) below. Each order of the perturbative
  expansion in powers of $\alpha_s(Q)$ contains up to two logarithms per
  loop of $Q\Tsc{b}$.  These logarithms are evidently infinite at
  $\Tsc{b}=0$, and fixed order perturbative calculations are entirely
  inapplicable to $\int \diff[2]{\T{q}} W$ with the original CSS
  definition.

  Recall that $W$ is an approximation to the cross section only for
  $\Tsc{q} \ll Q$.  Thus the transverse-coordinate-space quantity
  $\tilde{W}(\Tsc{b},Q)$ is important for a physical cross section only
  for $\Tsc{b}$ bigger than about $1/Q$.  Finite perturbative orders
  of the collinear expansion are useful when $\Tsc{b}$ is of order
  $1/Q$. 

\item Even without the issue of $W(\Tsc{q})$ becoming negative at large
  $\Tsc{q}$, there is the issue that it involves, in momentum space, a
  convolution of two independent TMD densities.  At large $\Tsc{q}$,
  these can be computed perturbatively in terms of collinear
  parton distribution functions (pdfs) and/or collinear fragmentation functions (ffs). 
  Power counting indicates that they are roughly of order
  $1/\Tsc{q}^2$.  Therefore, the basic TMD factorization formula gives
  a cross section that has this same power counting, and extends
  infinitely far beyond the kinematic limit.  The $Y$ term compensates
  this in principle, but the different perturbative truncations in $Y$
  and $W$ imply that the result can be numerically a bad
  approximation.
\end{itemize}
The culprit in each of the above is that the TMD factorization formula
used in $W(\Tsc{q})$ was derived to be a good approximation to the
cross section for $\Tsc{q} \ll Q$, but in the integral over $\T{q}$,
the formula is being used far beyond its domain of applicability.  

 There is a uniqueness to the particular form of $W(\Tsc{q})$ that
  gives rise to its undesirable properties at large $\Tsc{q}$.  The
  uniqueness arises from the use of a strict leading power expansion
  in $\Tsc{q}/Q$ when constructing the TMD factorization formula for
  $W$.  As an illustration, consider a lowest-order perturbative
  expansion that gives in $W$ a factor $\alpha_s \ln(Q/\Tsc{q}) /
  \Tsc{q}^2$ at small $\Tsc{q}$, with its characteristic logarithm.
  The use of exactly a single power of $\Tsc{q}$ (times logarithms)
  entails keeping the same formula at large $\Tsc{q}$, where the
  logarithm becomes negative.  

  The use of a strict leading power in $\Tsc{q}/Q$ is is important
  because the non-leading powers are much more complicated and often
  non-factorizing.  This issue is particularly important because, to
  leading power, gluons can connect subgraphs in different kinematic
  regions.  To get factorization, Ward identities are used to extract
  these gluons into attachments to Wilson lines in operator
  definitions of the correlation functions like TMD pdfs and ffs.
  However, the Ward identities apply only in the context of an
  approximation that is valid at leading power (or perhaps one power
  beyond).  The result is the afore-mentioned uniqueness in the
  factorized form. Essentially the same considerations apply in SCET
  for essentially the same reasons --- see
  Ref.~\cite{Rothstein:2016bsq}.

  It therefore becomes quite non-trivial to adjust the TMD
  factorization formula to get nicer properties at large $\Tsc{q}$
  without violating the derivation of TMD factorization.

Many implementations of TMD factorization calculate TMD functions by effectively resuming logarithms of $\Tsc{b} Q$. 
The usefulness this type of resummation assumes that there is a broad range of $\Tsc{b}$ where $1/Q \ll \Tsc{b} \ll 1/m$. At smaller $Q$ the 
window satisfying this condition shrinks and eventually vanishes, so that the advantage of such techniques becomes questionable.  
Moreover, errors introduced by including the region where $\Tsc{b} \ll 1/Q$
can start to become a significant fraction of the resummation calculation.

The situation is simpler if one simply works in a leading logarithm
approximation as in the work of Parisi and Petronzio (PP)
\cite{Parisi:1979se}.  There an ad hoc modification to impose rough
approximations to the true kinematics is appropriate.  But
modifications are much harder to impose in the middle of a full proof
of factorization that is to be applied generally.

Our approach in this paper is to preserve the factorized form of
$\tilde{W}(\Tsc{b})$ in transverse coordinate space, but to modify the
way in which it is used to construct a contribution from $W(\Tsc{q})$ to the cross
section, to try to evade the problems listed above.  We must
preserve the property that $W$ gives a good approximation
to the cross section at low transverse momentum, including the
important region where $\Tsc{q}$ is in the non-perturbative region of
order $m$.  Naturally, the definition of $Y$ must be correspondingly
modified.

The paper is organized as follows: We provide a general background of the main issues 
in Sec.~\ref{sec:principles}, and outline the principles that will guide our matching procedure. 
We review the basic logic of the $W+Y$ method in Sec.~\ref{sec:largesmall}, and include some clarifying remarks.
Since an important component of our procedure is that it leaves the treatment of the $W$-term largely unaltered, 
we will also need to review the standard factorization and evolution of the $W$-term in the CSS TMD factorization formalism, which we
do in Sec.~\ref{sec:review}. Next, we will explain our modifications,
starting in Sec.~\ref{sec:bstarmod} with a modified treatment of the
standard $\bstarsc$-prescription. This will allow us to construct a
generalized $W$-term. From this we will obtain a correspondingly generalized 
$Y$-term in Sec.~\ref{sec:Yterm}. Thus we will have constructed a new $W+Y$ method, but with additional parameters. In Sec.~\ref{sec:together} we 
discuss how the principles from Sec.~\ref{sec:principles} constrain parametrizations. In Sec.~\ref{sec:bcfg}, we elaborate on 
technical steps needed to calculate in the new $W+Y$ prescription, and in Sec.~\ref{sec:demo} we demonstrate the utility of our treatment by calculating
the $Y$ term with simple parametrizations of collinear quark pdfs and ffs.
We conclude by summarizing our logic and commenting on ways forward in Sec.~\ref{sec:con}.

\section{Guiding Principles}
\label{sec:principles} 

The standard $W+Y$ construction relies on the fact that, at very large $Q$, 
there is a broad range where $m/\Tsc{q}$ and $\Tsc{q}/Q$ are 
both good small expansion parameters.
We suggest the following principles to guide the choice of an
  improved formalism:
\begin{enumerate}
\item When the $W$ term is integrated over all $\T{q}$, it should
    obey an ordinary collinear factorization property.  This implies
    that when the scales in the pdfs and ffs are set to $\mu = Q$, the
    result should agree with the ordinary factorization calculation
    for the integrated cross section to zeroth order in $\alpha_s(Q)$,
    thereby matching the parton-model result appropriately.  
\item For $\Tsc{q} \gtrsim O(Q)$, the cross section
 given by $W+Y$ should appropriately match fixed order collinear perturbation 
theory calculations for large transverse momentum. 
\item For very large $Q$, the normal $W + Y$ construction should
  automatically be recovered for the $m \ll \Tsc{q} \ll Q$ region, to leading power in $Q$.
\item The modified $W$ term should be expressed in terms of the
    same coordinate space quantity $\tilde{W}$ as before, in order
    that operator definitions of the pdfs and ffs can be used,
    together with their evolution equations.
\item The sum $W+Y$ should give a leading power approximation to the cross
    section over the whole range of $\Tsc{q}$. Fixed order expansions
    of $Y$ in collinear perturbation theory are suitable for
    calculating $Y$, while the usual solution of evolution equations
    is used for $W$.
\end{enumerate}
We will use these principles to strongly motivate our new constructions of $W$ and $Y$.

We emphasize here that many of the elements of this article have already 
been used in the past in various forms. Our purpose in this paper is to synthesize and systematize them.

For example, a detailed discussion of large and small $\Tsc{q}$ matching and the associated perturbation theory errors in intermediate 
regions of $\Tsc{q}$ appears in Ref.~\cite{Arnold:1990yk} -- see especially Sections 1.2-1.4 for a clear discussion.
The work of Catani-Trentadue-Turnock-Webber and related 
treatments, especially Bozzi-Catani-de Florian-Grazzini (BCFG) in~\cite{Bozzi:2005wk} 
replace $\ln (Q^2 \Tsc{b}^2)$ terms in a resummation with $\ln (Q^2 \Tsc{b}^2 + 1)$, thus cutting off the $\Tsc{b} \ll 1/Q$ contribution.  
This is similar to work by Parisi and Petronzio~\cite{Parisi:1979se} that used this method to handle the 
$\Tsc{b} \ll 1/Q$ region in a leading-log approach. 
BCFG also impose constraints on the relationship between integrated and transverse momentum dependent cross sections that are
very similar to our points 1) through 3) above.

Nadolsky, Stump and Yuan (NSY)~\cite{Nadolsky:1999kb} performed a CSS-style analysis of semi-inclusive 
deep inelastic scattering (SIDIS), but modified the large $\Tsc{q}$ behavior of their resummed term 
by introducing $\Tsc{q}/Q$ corrections to the $x$ and $z$ kinematic variables. 
Specifically, NSY modified the $W$-term at larger values of $\Tsc{q}/Q$ 
to improve matching asymptotic term as order $\Tsc{q}/Q$ corrections start to become large. 
By examining the kinematics of the process, they found that 
an improved matching is achieved if one replaces the standard $x$ and $z$ variables in the collinear pdfs and ffs of the $W$ term by\footnote{See the discussion regarding matching in Section VA of Ref.~\cite{Nadolsky:1999kb} and the 
comparison between the modified and unmodified treatments in Fig.~9 of Ref.~\cite{Nadolsky:1999kb}.} 
\begin{align}
x {}& \to \tilde{x} = x \left( \frac{\Tsc{q}^2 + Q^2}{Q^2} \right) \, , \label{eq:xtilde}  \\
z {}& \to \tilde{z}= z \left( \frac{\Tsc{q}^2 + Q^2}{Q^2} \right) \, . \label{eq:ztilde}
\end{align}

In Ref.~\cite[Eq.~(13.75)]{Collins:2011qcdbook}, Collins proposed to impose a direct cutoff on the large $\Tsc{q}$ part of the $W$-term.
Our method follows a very similar approach (see Sec.~\ref{sec:wterm}), with our $\Xi$ function in Eq.~\eqref{eq:Wnew} corresponding to 
Collins's $F(\Tsc{q}/Q)$, and our $\TTnew{}{}$ corresponding roughly to Collins's $L_F$.
Likewise, CSS introduced a mass-scale $Q_T^{\rm min} \sim m$ in Ref.~\cite{Collins:1984kg} to regulate the 
low $\Tsc{q}$ part of the $Y$-term calculation. The role of $Q_T^{\rm min}$ is analogous to what we will call $\lambda$ in Sec.~\ref{sec:largesmall}. 
The replacements in Eqs.~\eqref{eq:xtilde}--\eqref{eq:ztilde} are physically motivated in that they approximate the kinematic corrections 
on $x$ and $z$ momentum fractions that begin to be important at larger $\Tsc{q}$. See also Sec.~2.6 of Ref.~\cite{Guzzi:2013aja} for a review of the kinematical rescaling procedure.

In most implementations of the ResBos Monte Carlo, for both Drell-Yan and SIDIS, 
the computational algorithm automatically forces a switch between the $W$-term (there called the ``resummed term'') to a calculation done using purely fixed order perturbative QCD above 
some $\Tsc{q}$. In fact, this is useful also for improving the efficiency of computer calculations 
since it means that computationally intensive calculations of the $W$-term can be short circuited above some $\Tsc{q}$ 
without compromising the accuracy of the calculation. (See Refs.~\cite{Balazs:1997xd,Nadolsky:1999kb}.) For very low $\Tsc{q}$, the ResBos 
Monte Carlo switches off the $Y$-term for $\Tsc{q} \lesssim  0.5-1.0$~GeV~\cite{pavelprivate}.

Boer and den Dunnen~\cite{Boer:2014tka,Boer:2015uqa} used a method similar to BCFG, but implemented the transition 
to very small $\Tsc{b}$ by using a modified renormalization group scale (called $\mub'$).
This aspect of the Boer-den Dunnen approach is very similar to what we will use in this article.

We suggest that, to maintain context, it will be useful to read the articles listed above concurrently 
with this paper.

\section{$W$ and $Y$ Terms}
\label{sec:largesmall}

We start by reviewing the $W + Y$ construction. This will establish notational conventions to be used throughout the 
paper in addition to clarifying the logic of the $W + Y$ method. We will also introduce 
one of our modifications.

Consider a generic transverse momentum dependent cross section
that depends on a hard scale $Q$ and is differential in a transverse
momentum $q_T$. It may also be differential in other kinematical
variables, but for simplicity we will not show these explicitly.  
It could be any cross section for which a TMD
factorization theorem exists. We will use the abbreviated notation
\begin{equation}
\label{eq:firstequation}
\cs{}{} = \frac{\diff{} \sigma}{\diff[2]{\T{q}} \diff{Q} \cdots} \,\,  .
\end{equation}
The ellipsis indicates possible dependence on other kinematical
variables like $z$ and $x$, whose exact values are not relevant to our
immediate discussion. Although the logic in this paper is meant to apply generally,  explicit expressions
will be written for SIDIS. CSS-style derivations of TMD factorization are 
given for SIDIS in Refs.~\cite{Meng:1991da,Meng:1995yn} (see also~\cite[Sec.~13.15]{Collins:2011qcdbook}). 

The TMD formalism separates Eq.~\eqref{eq:firstequation} into a sum of
two terms. One term ($W$) describes the small transverse momentum
behavior $\Tsc{q} \ll Q$ and an additive correction term ($Y$)
accounts for behavior at $\Tsc{q} \sim Q$:
\begin{equation}
\cs{}{} = \TT{}{} + \YY{}{} + O\xleft( \frac{m}{Q} \right)^c \cs{}{} \, . \label{eq:basic}
\end{equation}
The first term on the right is written in terms of TMD pdfs and/or TMD ffs and is constructed to be an accurate description in the limit of
$\Tsc{q}/Q \ll 1$.  It includes all 
non-perturbative transverse momentum dependence.
The $Y$-term is described entirely in terms of
\emph{collinear}
factorization.  Our aim is to construct $W$ and $Y$ such that
  $W+Y$ gives the cross section up to an error that, relative to the
  cross section, is of order
a positive ($c>0$) power of $m/Q$, where $m$ is a
hadronic mass scale. 

The original CSS definition of $W$ is as given in, for
example, Ref.~\cite[13.71]{Collins:2011qcdbook} (where it is called
$L$):
\begin{equation}
\label{eq:wterm}
\TT{}{} \equiv \appor{TMD} \cs{}{}  \,. 
\end{equation}
The $\appor{TMD}$ ``approximator'' is an instruction to replace the object
to its right by an approximation that is designed to be good in the
$\Tsc{q} \ll Q$ limit. That is, it replaces the exact $\cs{}{}$ by the
approximate $\TT{}{}$:
\begin{align}
 \appor{TMD} \cs{}{} = \cs{}{} 
          & + O \xleft( \frac{\Tsc{q}}{Q} \right)^a \cs{}{}
\nonumber \\ 
      &  + O \xleft( \frac{m}{Q} \right)^{a'} \cs{}{}
\, ,
\label{eq:TMDapdef}
\end{align} 
where $a, a' >0$.

Another approximator, $\appor{coll}$, handles the
large $\Tsc{q} \sim Q$ region. It replaces $\cs{}{}$ with an
approximation that is good when $\Tsc{q} \sim Q$. That is, 
\begin{align}
\appor{coll} \cs{}{} = \cs{}{} 
      & + O \xleft( \frac{m}{\Tsc{q}} \right)^b \cs{}{}
 \, , \label{eq:collapdef}
\end{align} 
where $b>0$.
Since $\appor{coll}$ is to be applied to the
$\Tsc{q} \sim Q$ region, one only needs collinear factorization
at a fixed order and with a hard scale $\mu \sim Q$.

If $\Tsc{q} \lesssim m$ and $\Tsc{q} \sim Q$ were the only regions of
interest, then the $\appor{TMD}$ and $\appor{coll}$ approximators would be sufficient. One could
simply calculate using fixed order collinear factorization for the
large $\Tsc{q}$-dependence and TMD factorization for small $\Tsc{q}$-dependence.  
A reasonable description of the full transverse momentum
dependence would be obtained by simply interpolating between the
two descriptions~\cite{Chay:1991jc,Anselmino:2006rv}.

However, the region between large and small $\Tsc{q}$ needs special
treatment if errors are to be strictly power suppressed point-by-point
in $\Tsc{q}$.  The standard method is to construct a sequence of
nested subtractions. The smallest-size region is a neighborhood of
$\Tsc{q} = 0$, where $\appor{TMD}$ gives a very good approximation.
So, one starts by adding and subtracting the $\appor{TMD}$
approximation:
\begin{align}  \cs{}{} \, =  \, & \appor{TMD} \cs{}{}  \no
& \;\;  + \Bigg[ \cs{}{} - \appor{TMD} \cs{}{} \Bigg] \, .
\label{eq:nextapp}
\end{align}
From Eq.~\eqref{eq:TMDapdef}, the error term in the square brackets is order $( \Tsc{q}/Q )^a$ and is
only unsuppressed at $\Tsc{q} \gg m$.
Therefore, one may apply $\appor{coll}$ and then use a fixed-order
  perturbative expansion in collinear factorization:
\begin{align}
\Gamma( & m \lesssim \Tsc{q} \lesssim Q,Q) 
\nonumber\\
={}&   \appor{TMD} \cs{}{}
 + \appor{coll} \left[ \cs{}{} - \appor{TMD} \cs{}{} \right]
\nonumber\\
&
 + O\xleft( \left( \frac{m}{\Tsc{q}} \right)^b \left( \frac{\Tsc{q}}{Q} \right)^a  \right) \cs{}{}
\no
&
 + O\xleft( \left( \frac{m}{\Tsc{q}} \right)^b \left( \frac{m}{Q} \right)^{a'}  \right) \cs{}{} 
\no
={}& \TT{}{} +  \appor{coll}\cs{}{} -  \appor{coll}\appor{TMD} \cs{}{}
\no 
& + O\xleft(  \frac{m}{Q}\right)^{\rm c} \cs{}{}
\, ,
\label{eq:powercounting}
\end{align}
where $c = \min(a,a',b)$.  Thus, the cross section is determined
point-by-point in the mid-$\Tsc{q}$ region, up to powers of $m/Q$, by a combination of TMD and
collinear correlation functions. 

The CSS construction of $W+Y$ defines $W$ and $Y$ to be the
  first and second terms on the second line of Eq.\
  (\ref{eq:powercounting}).  Their specific definitions of
  $\appor{coll}$ and $\appor{TMD}$ allowed Eq.\
  (\ref{eq:powercounting}) to work only in the $m
  \lesssim \Tsc{q} \lesssim Q$ region, which we emphasize by the
  argument on the left side of Eq.~\eqref{eq:powercounting}.  The
  error estimates in Eq.~\eqref{eq:powercounting} are inapplicable
  outside this range, i.e., they must not be applied when $\Tsc{q} \gg
  Q$ or $\Tsc{q} \ll m$.  This is because they were extracted from the
  leading power of expansions in relatively small kinematic variables
  $\Tsc{q}/Q$ and $m/\Tsc{q}$
  to give Eqs.~(\ref{eq:TMDapdef}) and~(\ref{eq:collapdef}).
  The issues are illustrated by Eq.\ (\ref{eq:collapdef}).  The
  $(m/\Tsc{q})^b$ estimate is obtained from an expansion in powers of
  mass with respect to the smallest scale in the collinear
  hard-scattering; it is of the order of the first omitted term in the
  expansion.  But once $\Tsc{q}$ gets much smaller, the error can be
  arbitrarily larger.  As a mathematical example, suppose
  \begin{equation}
    \Gamma = \frac{1}{ (\Tsc{q}^2 + m^2)^2 }.
  \end{equation}
  The leading power expansion in $m/\Tsc{q}$ is
  \begin{equation}
    \appor{coll} \Gamma = \frac{1}{ \Tsc{q}^4 },
  \end{equation}
  and the error is 
  \begin{equation}
    \Gamma -\appor{coll} \Gamma
     = \left( - \frac{2m^2}{ \Tsc{q}^2 } - \frac{m^4}{ \Tsc{q}^4 }
       \right) \Gamma.
  \end{equation}
  For the error estimate when $m \lesssim \Tsc{q}$, we can correctly
  take $b=2$:
  \begin{equation}
    \Gamma -\appor{coll} \Gamma
     = O\xleft( \frac{m^2}{ \Tsc{q}^2 } \right) \Gamma.
  \end{equation}
  But when $\Tsc{q} \ll m$, the error is a stronger behavior,
  $m^4/\Tsc{q}^4$ relative to $\Gamma$.

  It is useful to review the precise meaning of notation in the
  error estimates, which is as follows: An $O(\Tsc{q}/Q)$ error means
  that there exist constant positive real numbers, $\mathcal{C}$ and
  $\mathcal{A}$, such that the error is less than $\mathcal{C}
  \Tsc{q}/Q$ for $\Tsc{q}/Q < \mathcal{A}$. Analogous statements apply
  to $O(m/\Tsc{q})$ and $O(m/Q)$ error estimates. Thus, the error
  estimates in Eqs.~\eqref{eq:basic}--\eqref{eq:powercounting} provide
  no constraints on the behavior in the $\Tsc{q} \gtrsim Q$ or
  $\Tsc{q} \lesssim m$ regions.  As shown above, the true errors in
  those regions could be much worse than a naive extrapolation of the
  powers in Eqs.~\eqref{eq:basic}--\eqref{eq:powercounting} would
  suggest.

The above observations do not represent a fundamental breakdown of the
formalism.  They merely indicate that some extra care is needed to
  construct a formalism valid also for 
  $\Tsc{q} \lesssim m$ and $\Tsc{q} \gtrsim Q$.

For $\Tsc{q} \lesssim m$, collinear factorization is
certainly not applicable for the differential cross section.  But
  this region is actually where the $W$-term in
  Eq.~\eqref{eq:TMDapdef} has its highest validity.  So one simply
  must ensure that the would-be $Y$-term
\begin{equation}
\appor{coll} \cs{}{} - \appor{coll} \appor{TMD} \cs{}{}
\end{equation}
  is sufficiently suppressed in Eq.~\eqref{eq:powercounting} for
  $\Tsc{q} \lesssim m$.  Therefore, we will modify the usual
  definition of $Y$ by inserting a suppression factor at low
  $\Tsc{q}$:
\begin{align}
\label{eq:yterm}
& \YY{}{}  \nonumber \\ 
&{}\equiv \left\{ \appor{coll} \left[ \cs{}{} - \TT{}{} \right] \right\} X(\Tsc{q}/\lambda) \nonumber \\
   &{}= \left\{ \appor{coll} \cs{}{} - \appor{coll} \appor{TMD} \cs{}{} \right\} X(\Tsc{q}/\lambda) \, . 
\end{align}
The smooth cutoff
function $X(\Tsc{q}/\lambda)$  approaches zero for $\Tsc{q}
\lesssim \lambda$ and unity for $\Tsc{q} \gtrsim \lambda$.  It ensures
that the $Y$-term is a correction for $\Tsc{q} \gtrsim m$ only.  As
long as $\lambda = O(m)$, any $\lambda$-dependence must be weak.
This is analogous to the introduction of a $Q_T^{\rm min}$ in Ref.~\cite[Eq.~(2.8)]{Collins:1984kg}.

The exact functional form of $X(\Tsc{q}/\lambda)$ is arbitrary, but is most useful in calculations if it sharply 
suppresses $\Tsc{q} \ll m$ contributions while not affecting $\Tsc{q} \gtrsim m$.  While a step function is acceptable,  
we suggest using a slightly smoother function since one expects the transition from perturbative to non-perturbative physics to 
be relatively smooth. One possible choice is
\begin{equation}
X(\Tsc{q}/\lambda) = 1 - \exp \left\{ -(\Tsc{q} / \lambda)^{a_X} \right\} \ . \label{eq:Xparam}
\end{equation}
This is what we will use in sample calculations in Sec.~\ref{sec:demo}. A large value for the 
power $a_X$ makes the switching function more like a step function. 

In common terminology, the first term in braces on the second line of Eq.~\eqref{eq:yterm} is
called the ``fixed order'' (FO) contribution, while the second term is 
the ``asymptotic'' (AY) contribution. We will
use the notation
\begin{align}
\fixo{}{} & \equiv \appor{coll} \cs{}{} \label{eq:fodef}  \\
\as{}{} &\equiv \appor{coll} \appor{TMD} \cs{}{}  \label{eq:asydef} \, .
\end{align}
So,
\begin{equation}
\YY{}{} \equiv \left\{ \fixo{}{} - \as{}{}  \right\} X(\Tsc{q}/\lambda) \, .
\label{eq:Y_}
\end{equation}
This corresponds to the terminology in, for example, Ref.~\cite{Nadolsky:1999kb}. The term ``fixed order'' is meant to imply 
that the calculation of $\Gamma$ is done entirely with collinear factorization with hard parts calculated to low order in perturbation theory using $\mu = Q$ and 
with collinear pdfs and ffs calculated using $\mu = Q$. That is, the hard part and the parton correlation functions are evaluated at the same scale.

Now we can extend the power
suppression error estimate in Eq.~\eqref{eq:powercounting} down to
$\Tsc{q} = 0$ to recover Eq.~\eqref{eq:basic}.  
Equation~\eqref{eq:powercounting} becomes
\begin{align}
\label{eq:basic2}
\Gamma(\Tsc{q} \lesssim Q,Q)  = &\TT{}{} +  \YY{}{}\no
& + O\xleft( \frac{m}{Q}\right)^{\rm c} \cs{}{},
\end{align} 
which is Eq.~\eqref{eq:basic}, but restricted to $\Tsc{q} \lesssim Q$.

So far, aside from introducing an explicit $X(\Tsc{q}/\lambda)$, we have only 
reviewed the standard $W+Y$ construction. The $\Tsc{q} \lesssim Q$ restriction on 
the left of Eq.~\eqref{eq:basic2} should be emphasized. Since we rely 
on strict power counting in $\Tsc{q}/Q$ and $m/\Tsc{q}$, the region of $\Tsc{q} \gtrsim Q$ is 
not guaranteed to be well-described by the above $W+Y$ construction. We will correct this in 
Secs.~\ref{sec:wterm}--\ref{sec:together}
with a modified $W$-term definition.

\section{Review of TMD Factorization and Basic Formulas}
\label{sec:review}

Our proposed modifications to the transition to the $\Tsc{q} / Q \gtrsim 1$ region will leave the  
standard treatment of TMD factorization~\cite[Chapters
10,13,14]{Collins:2011qcdbook} in the $\Tsc{q} / Q \ll 1$ region
only slightly modified.\footnote{See also Ref.~\cite{Rogers:2015sqa} for a recent brief overview and large list of references relating to the development of TMD factorization.} In particular, 
the operator definitions for transverse-coordinate-space TMD
functions, along with their evolution properties, are exactly the same as in the usual formalism. 
This is an important aspect of our suggested modifications, so it is worthwhile to review the basics 
of TMD factorization for the low $\Tsc{q}$ region. This section gives a short summary of the most important 
formulas, with the organization of notation optimized for discussions in later sections. We will also
refer frequently to the review of TMD evolution in Ref.~\cite[Sec.~II]{Collins:2014jpa}, especially~\cite[Eqs.~(22, 24)]{Collins:2014jpa}. 

\subsection{TMD Evolution} 
\label{sec:evolution}

The evolution of $\TT{}{}$ follows from generalized renormalization properties of the operator definitions 
for TMD pdfs and ffs. To separate perturbative and non-perturbative contributions, 
one defines large and small $\Tsc{b}$ through a function $\bstarsc$ that 
freezes above some $\bmax$ and equals $\Tsc{b}$ for small $\Tsc{b}$:
\begin{equation}
\bstarsc(\Tsc{b}) \longrightarrow
\begin{dcases}
\Tsc{b} & \Tsc{b} \ll \bmax \\
\bmax & \Tsc{b} \gg \bmax \, . \label{eq:bdef}
\end{dcases}
\end{equation}
The relevant renormalization group scales are
\begin{equation}
\label{eq:mubdef}
\mub \equiv C_1/\Tsc{b}  \, ,\qquad \mubstar \equiv C_1/\bstarsc \, , \qquad  \muQ \equiv C_2 Q \, ,
\end{equation}
where $C_1$ and $C_2$ are constants that are chosen to optimize perturbative convergence.

  We first solve the evolution equations, to give the following
  forms for the $W$-term for SIDIS (neutral-current and neglecting
  heavy flavors):
\begin{widetext}
\begin{align}
\TT{}{}
  =&{} \sum_{j} H_{j}(\mu_Q,Q)
    \int \frac{\diff[2]{\T{b}}}{(2 \pi)^2}
    e^{i\T{q}\cdot \T{b} }
    \tilde{F}_{j/A}\big( x_A, \T{b}  ; Q_0^2, \mu_{Q_0} \bigr)
    \,
    \tilde{D}_{B/j} \big( z_B, \T{b} ; Q_0^2, \mu_{Q_0} \bigr)
\nonumber\\&
 \, \times  \exp\left\{   
           \int_{\mu_{Q_0}}^{\muQ}  \frac{ \diff{\mu'} }{ \mu' }
           \biggl[ 2 \gamma(\alpha_s(\mu'); 1) 
                 - \ln\frac{Q^2}{ (\mu')^2 } \gamma_K(\alpha_s(\mu'))
           \biggr]
           + \tilde{K}(\Tsc{b};\mu_{Q_0})
             \ln \left( \frac{ Q^2 }{ Q_0^2 } \right)
  \right\} 
\nonumber\\
  =&{} \sum_{j} H_{j}(\mu_Q,Q)
    \int \frac{\diff[2]{\T{b}}}{(2 \pi)^2}
    e^{i\T{q}\cdot \T{b} }
    \tilde{F}_{j/A}\big( x_A, \T{b}  ; Q_0^2, \mu_{Q_0} \bigr)
    \,
    \tilde{D}_{B/j} \big( z_B, \T{b} ; Q_0^2, \mu_{Q_0} \bigr)
\nonumber\\&
 \, \times  \exp\left\{   
           \int_{\mu_{Q_0}}^{\muQ}  \frac{ \diff{\mu'} }{ \mu' }
           \biggl[ 2 \gamma(\alpha_s(\mu'); 1) 
                 - \ln\frac{Q^2}{ (\mu')^2 } \gamma_K(\alpha_s(\mu'))
           \biggr]
  \right\} \nonumber\\&
    \,\times
  \exp \left\{ \left[ \tilde{K}(\Tsc{b};\mubstar) - \int_{\mubstar}^{\mu_{Q_0}} \frac{d\mu'}{\mu'} \gamma_K(\alpha_s(\mu')) \right]  \ln \left( \frac{ Q^2 }{ Q_0^2 } \right) \right\}\,   .
\label{eq:solnf}
\end{align}
Here $\tilde{F}_{j/A}\big( x_A, \T{b}  ; Q_0^2, \mu_{Q_0} \bigr)$,
and $\tilde{D}_{B/j} \big( z_B, \T{b} ; Q_0^2, \mu_{Q_0} \bigr)$ are,
respectively, the TMD pdf and TMD ff evaluated at a reference scale
$Q_0$.
Their operator definitions are given in Eqs.~(13.42,13.106) of
Ref.~\cite{Collins:2011qcdbook}.  The exponential factor on the
  second line implements the evolution from {$Q_0$ to
  $Q$}. There $\tilde{K}(\Tsc{b};\mu)$ is the Collins-Soper (CS)
evolution kernel (see~\cite[Eq.~(6,11,25)]{Collins:2014jpa}), while
$\gamma_K(\alpha_s(\mu))$ and $\gamma(\alpha_s(\mu'); 1)$ are
anomalous dimensions for the CS kernel and a TMD pdf/ff respectively
(see~\cite[Eq.~(7,8,9,10,12)]{Collins:2014jpa}).  See also
Refs.~\cite{Rogers:2015sqa,Collins:2012ss} and references therein for
detailed discussions of the evolution equations and their origins.
In the last part of Eq.\ (\ref{eq:solnf}), we have used the
renormalization group to change the $\mu$ argument of $\tilde{K}$ from
$\mu_{Q_0}$ to $\mubstar$.  This is in anticipation of later
manipulations, where $\mubstar$ will be a suitable scale for
perturbatively calculated quantities.

We define the $\Tsc{b}$-space version $\tilde{W}$ of the $W$-term by
\begin{equation}
\label{eq:FTdef}
\TT{}{} = \int \frac{\diff[2]{\T{b}}}{(2 \pi)^2} e^{i\T{q}\cdot \T{b} } \, \tilde{W}(\Tsc{b},Q) \, .
\end{equation}
To economize notation, we will assume there is only one flavor of parton so that we may drop the 
sum over $j$ and the $j$
subscript. In detailed calculations, the sum needs to be restored.\footnote{Recall, however, that for scattering off a quark, there is no flavor dependence in the hard scattering until order $\alpha_s^3$. 
So flavor independence is likely a good approximation. See the discussion at the beginning of section VIA of Ref.~\cite{Collins:2014jpa}. }

In the limit $\Tsc{b} \ll 1/m$, each TMD correlation function can be expanded in an OPE and expressed 
in terms of collinear correlation functions. Then the transverse coordinate dependence is itself perturbatively generated. 
Let us define a notation to describe this limit. First, substitute 
$\Tsc{b} \to \bstarsc$ in Eq.~\eqref{eq:solnf} to regulate the 
$\Tsc{b} \gtrsim 1/m$ region. Second, expand the result in an OPE and
drop order $O(\Tsc{b} m)$ corrections.  Finally we replace $Q_0$ and
$\mu_{Q_0}$ by $\mubstar$, so that perturbatively calculations have no
large logarithms.
We call 
the result $\tilde{W}^{\rm OPE}(\bstarsc(\Tsc{b}),Q)$:
\begin{align}
\tilde{W}^{\rm OPE}(\bstarsc(\Tsc{b}),Q)  
  \equiv{} & H(\mu_{Q},Q) \sum_{j' i'} \int_{x_A}^1
\frac{d \hat{x}}{\hat{x}}  \tilde{C}^{\rm pdf}_{j/{j'}}(x_A/\hat{x},\bstarsc(\Tsc{b});\mubstar^2,\mubstar,\alpha_s(\mubstar)) f_{j'/A}(\hat{x};\mubstar) \times \nonumber \\
 & \times  \int_{z_B}^1 \frac{d \hat{z}}{\hat{z}^3} \tilde{C}^{\rm ff}_{i'/{j}}(z_B/\hat{z},\bstarsc(\Tsc{b});\mubstar^2,\mubstar,\alpha_s(\mubstar)) d_{B/i'}(\hat{z};\mubstar) \times \nonumber \\
& \times \exp \left\{ \ln \frac{Q^2}{\mubstar^2} \tilde{K}(\bstarsc(\Tsc{b});\mubstar) + 
\int_{\mubstar}^{\mu_{Q}} \frac{d \mu^\prime}{\mu^\prime} \left[ 2 \gamma(\alpha_s(\mu^\prime);1) 
- \ln \frac{Q^2 }{{\mu^\prime}^2} \gamma_K(\alpha_s(\mu^\prime)) \right]\right\} \, . \label{eq:Tcoll}
\end{align}
The functions $f_{{j'}/A}(x;\mu)$ and
 $d_{B/{j'}}(z;\mu)$ are the ordinary collinear pdf and ff. 
Equation~\eqref{eq:Tcoll} is the standard result for the small $\Tsc{b}$ limit and corresponds to Eq.~(22) of Ref.~\cite{Collins:2014jpa}, but without 
the non-perturbative exponential factors.
Thus, 
\begin{equation}
\tilde{W}(\Tsc{b},Q) = \tilde{W}^{\rm OPE}(\bstarsc(\Tsc{b}),Q) + O\xleft( ( \Tsc{b} m )^p \right) \,  \label{eq:smallblimit}
\end{equation}
with $p > 0$. 

\subsection{Separation of Large and Small $\Tsc{b}$}
\label{sec:organization}

For Eq.~\eqref{eq:bdef}, a common functional form is~\cite{Collins:1981va}:
\begin{equation}
\label{eq:bstardeff}
\bstarsc(\Tsc{b}) \equiv  \sqrt{\frac{\Tsc{b}^2}{1 + \Tsc{b}^2 / \bmax^2}} \, .
\end{equation}
The standard steps for separating large and small $\Tsc{b}$ are to first write a ratio,
\begin{equation}
  \label{eq:gjH.def0}
  e^{-g_{A}(x_A,\Tsc{b};\bmax)-g_{B}(z_B,\Tsc{b};\bmax)} 
  \equiv \frac{ \tilde{W}(\Tsc{b},Q_0) }
         { \tilde{W}^{\rm OPE}(\bstarsc(\Tsc{b}),Q_0) } \, .
\end{equation}
The ratio on the right side \emph{defines} the exponential functions on the left according to some reference scale $Q_0$. 
The $g$-functions, therefore, account for all the error terms on the right side~\eqref{eq:smallblimit} (at some $Q_0$).\footnote{It is essentially just convention that the 
$g$-functions appear in an exponent.} 
Next, one notices that the CS evolution is identical for the numerator and denominator, apart 
from the fact that the evolution kernel is evaluated at $\Tsc{b}$ in the former and $\bstarsc(\Tsc{b})$ in the latter. 
Thus, one may re-express the right side of Eq.~\eqref{eq:gjH.def0} in terms of $\tilde{W}$ at an arbitrary $Q$ in a very simple form by applying CS evolution 
to the numerator and denominator separately and canceling out many common evolution factors. The result is
\begin{equation}
  \label{eq:gjH.def}
  e^{-g_{A}(x_A,\Tsc{b};\bmax)-g_{B}(z_B,\Tsc{b};\bmax)} 
  = \frac{ \tilde{W}(\Tsc{b},Q) }
         { \tilde{W}^{\rm OPE}(\bstarsc(\Tsc{b}),Q) }
    e^{2 g_K(\Tsc{b};\bmax) \ln(Q/Q_0)} \, .
\end{equation}
Here, $g_K(\Tsc{b};\bmax)$
is the difference between the CS
evolution kernels evaluated at $\Tsc{b}$ and $\bstarsc(\Tsc{b})$:
\begin{equation}
\label{eq:gK.def}
g_K(\Tsc{b};\bmax) \equiv -K(\Tsc{b},\mu)+K(\bstarsc(\Tsc{b}),\mu) \, .
\end{equation} 
Now the kernel $\tilde{K}(\Tsc{b};\mu)$ is very strongly
universal; it is independent not just of the process, but also of
scale, polarization, $x$, $z$, flavors, and polarization.  The
``non-perturbative'' function $g_K(\Tsc{b};\bmax)$, defined 
by Eq.\ (\ref{eq:gK.def}), inherits the same strong universality properties as
$K(\Tsc{b},\mu)$.

Equation~\eqref{eq:gjH.def} allows us to write
\begin{equation}
\TT{}{}
  ={} 
    \int \frac{\diff[2]{\Tsc{b}}}{(2 \pi)^2}
    e^{i\T{q}\cdot \Tsc{b} }  \tilde{W}^{\rm OPE}(\bstarsc(\Tsc{b}),Q) \tilde{W}_{\rm NP}(\Tsc{b},Q;\bmax) \label{eq:solngb} \, , 
\end{equation}
where $\tilde{W}_{\rm NP}(\Tsc{b},Q;\bmax)$ is the combination 
of all non-perturbative exponential functions in Eq.~\eqref{eq:gjH.def},
\begin{align}
\tilde{W}_{\rm NP}(\Tsc{b},Q;\bmax)= e^{-g_{A}(x_A,\Tsc{b};\bmax)-g_{B}(z_B,\Tsc{b};\bmax)}\, 
e^{-2 g_K(\Tsc{b};\bmax) \ln(Q/Q_0)}\, . \label{eq:npparts}
\end{align}
$\tilde{W}_{\rm NP}(\Tsc{b},Q;\bmax)$ is a function to be
parameterized and fit to data, or to be determined by
appealing to non-perturbative methods.\footnote{To call $g_A$,
    $g_B$, and $g_K$ functions ``non-perturbative'' is somewhat
    of a misnomer.  The definition of $\tilde{W}^{\rm NP}$ is indeed
    such that it does include all the strongly non-perturbative
    contributions.  But if $\bmax$ is conservatively small,
    $\tilde{W}^{\rm NP}$ also includes contributions, at moderate
    $\Tsc{b}$,  that could be estimated perturbatively.}
$\tilde{W}^{\rm OPE}(\bstarsc(\Tsc{b}),Q)$ is calculable in collinear
factorization in terms of collinear pdfs and ffs and allows the use of low order perturbation 
theory for perturbatively calculable parts. It is exactly the original
definition of $\tilde{W}$, but evaluated at $\bstarsc(\Tsc{b})$ instead of $\Tsc{b}$.  
The exponential factors in Eq.~\eqref{eq:npparts} account for the non-perturbative 
transverse coordinate dependence. Notice that by construction
\begin{equation}
\frac{\diff{} }{\diff{\bmax}} 
\left[ 
   \tilde{W}^{\rm OPE}(\bstarsc(\Tsc{b}),Q)
   \tilde{W}_{\rm NP}(\Tsc{b},Q;\bmax) 
\right]
= 0 
\, .
\end{equation}

Substituting Eqs.~(\ref{eq:Tcoll}) and (\ref{eq:npparts}) into Eq.~\eqref{eq:solngb} produces 
the most familiar representation of the evolved $\tilde{W}(\Tsc{b},Q)$:
\begin{eqnarray}
\tilde{W}(\Tsc{b},Q)  
 & = & H(\mu_{Q},Q) \sum_{j' i'} \int_{x_A}^1
\frac{d \hat{x}}{\hat{x}}  \tilde{C}^{\rm pdf}_{j/{j'}}(x_A/\hat{x},\bstarsc(\Tsc{b});\mubstar^2,\mubstar,\alpha_s(\mubstar)) f_{j'/A}(\hat{x};\mubstar) \times \nonumber \\
& & \times  \int_{z_B}^1 \frac{d \hat{z}}{\hat{z}^3} \tilde{C}^{\rm ff}_{i'/{j}}(z_B/\hat{z},\bstarsc(\Tsc{b});\mubstar^2,\mubstar,\alpha_s(\mubstar)) d_{B/i'}(\hat{z};\mubstar) \times \nonumber \\
& \times & \exp \left\{ \ln \frac{Q^2}{\mubstar^2} \tilde{K}(\bstarsc(\Tsc{b});\mubstar) + 
\int_{\mubstar}^{\mu_{Q}} \frac{d \mu^\prime}{\mu^\prime} \left[ 2 \gamma(\alpha_s(\mu^\prime);1) 
- \ln \frac{Q^2 }{{\mu^\prime}^2} \gamma_K(\alpha_s(\mu^\prime)) \right]\right\}  \nonumber \\ 
 & \times &  \exp\left\{ -g_{A}(x_A,\Tsc{b};\bmax)-g_{B}(z_B,\Tsc{b};\bmax)  
-2 g_K(\Tsc{b};\bmax) \ln \left( \frac{Q}{Q_0} \right) \right\} \, . 
\label{eq:finalevolved}
\end{eqnarray}
This now includes all the necessary non-perturbative functions and corresponds to Eq.~(22) of Ref.~\cite{Collins:2014jpa}. In the case that 
non-perturbative functions are dropped, the $W$-term matches Eq.~(1.1) of Ref.~\cite{Collins:1984kg}.
 
With the method of Eqs.~\eqref{eq:bstardeff}--\eqref{eq:finalevolved}, the relationship between $\tilde{W}^{\rm OPE}(\bstarsc(\Tsc{b}),Q)$ and $ \tilde{W}_{\rm NP}(\Tsc{b},Q;\bmax)$ and 
the exact definition of $\tilde{W}$ from the factorization derivation is kept explicit. 
Equation~\eqref{eq:gjH.def} is exact because the evolution is the same for the numerator and denominator. Therefore, all $O\xleft( ( \Tsc{b} m )^p \right)$ corrections in 
Eq.~\eqref{eq:smallblimit} are accounted for automatically in the definition of the non-perturbative parts in Eq.~\eqref{eq:npparts}. The only errors in the relationship 
between the $W$-term and the physical cross section are the overall $m/Q, \, \Tsc{q}/Q$-suppressed errors from the factorization derivation.  

This section has been a compressed review of steps already reviewed recently in Sec.~2.B.III of Ref.~\cite{Collins:2014jpa}. We refer the 
reader to this and references therein for more details.  
 

\end{widetext}

\section{Modified $\bstarsc$-prescription and $W$-Term}
\label{sec:bstarmod}
\label{sec:wterm}

  Next, we modify the definition of $W$.  This is to provide a
  convenient solution to the problem that with the definitions given
  so far, the integral over all $\T{q}$ of $W(\Tsc{q})$ is zero, because
  $\tilde{W}(\Tsc{b})$ is zero at $\Tsc{b}=0$ (see App.\ \ref{sec:Wzero}).

  It would be preferable for the integral to have a normal collinear
  expansion in terms of pdfs and ffs at scale $\muQ$; the lowest order
  term then reproduces the lowest order collinear factorization result
  for the integrated cross section.  At the same time, we wish to
  preserve the results for the $\T{b}$-space quantity
  $\tilde{W}(\Tsc{b})$, since these embody the derived factorization
  and evolution properties.
  Most importantly, the modified $W$ term must still approximate the
  cross section at low $\Tsc{q}$ to the same accuracy as in Eq.\
  (\ref{eq:TMDapdef}).

  We achieve the modified $W$ in two stages.

  The first is to modify the Fourier transform in Eq.\
  (\ref{eq:FTdef}) to read
\begin{equation}
\label{eq:FTdef1}
  \TTa{}{}
  =
   \int \frac{\diff[2]{\T{b}}}{(2 \pi)^2}
   e^{i\T{q}\cdot \T{b} } \, \tilde{W}(\bone(\Tsc{b}),Q) \, .
\end{equation}
where
\begin{equation}
  \bone(\Tsc{b}) = \sqrt{ \Tsc{b}^2 + b_0^2/(C_5Q)^2 } \, .
\label{eq:bcut}
\end{equation}
That is, $\tilde{W}(\Tsc{b},Q)$ is replaced by
$\tilde{W}(\bone(\Tsc{b}),Q)$.
 The function $\bone(\Tsc{b})$ is arranged to agree with $\Tsc{b}$
  when $\Tsc{b} \gg 1/Q$, but to be of order $1/Q$ when $\Tsc{b}=0$, ,
  thereby providing a cutoff at small $\Tsc{b}$.
Then, when (\ref{eq:FTdef1}) is integrated over $\T{q}$, we
get $\tilde{W}(b_0/(C_5Q),Q)$, instead of the previous value
$\tilde{W}(0,Q)=0$.  We have included an explicit numerical factor of $b_0 \equiv 2
\exp(-\gamma_E)$ since this tends to lead to simpler formulas later on.
We have chosen the value of $\bone(0)$ to be
proportional to $1/Q$, so that, from Eq.\ (\ref{eq:finalevolved}),
$\tilde{W}(b_0/(C_5Q),Q)$ has a normal collinear factorization
  property. The numerical constant $C_5$ fixes 
the exact proportionality between $\bone(0)$ and $1/Q$.

But at the same time (\ref{eq:FTdef1}) still gives an approximation to the cross section of
the appropriate accuracy.  This is because, when
$\Tsc{q} \ll Q$, the dominant range of $\Tsc{b}$ is much larger than
$1/Q$, and so the modification in (\ref{eq:FTdef1}) only gives a
power-suppressed contribution.  Of course, at large $\Tsc{q}$, there
are more substantial changes.  But then we approach the domain of
validity of collinear factorization, and so the accuracy of the $W+Y$ form
is preserved provided that, in the definition \eqref{eq:yterm} of $Y$,
we replace $\TT{}{}$ by $\TTa{}{}$.

Note that the integrand in (\ref{eq:FTdef1}) is non-singular at
$\Tsc{b}=0$, unlike (\ref{eq:FTdef}).  Thus the large $\Tsc{q}$
behavior is exponentially damped.  Even so, the function still extends
to arbitrarily large $\Tsc{q}$.  

So the second and final stage of modification for $W$ is to make an
explicit cutoff at large $\Tsc{q}$, to give:
\begin{multline}
  \label{eq:Wnew}
  \TTnew{}{}
\\
  \equiv
  \Xi\xleft(\frac{\Tsc{q}}{Q},\eta\right)
  \int \frac{\diff[2]{\T{b}}}{(2 \pi)^2}
    e^{i\T{q}\cdot \T{b} } \tilde{W}(\bone(\Tsc{b}),Q) \, .
\end{multline}
Here $\Xi\xleft(\Tsc{q}/ (Q\eta)\right)$ is a cutoff function that we
introduce to ensure that $\TTnew{}{}$ vanishes for $\Tsc{q} \gtrsim
Q$, and $\eta$ is a parameter to control exactly where the suppression
of large $\Tsc{q}$ begins. $\Xi\xleft(\Tsc{q}/Q,\eta\right)$ should
approach unity when $\Tsc{q}\ll Q$ and should vanish for $\Tsc{q}
\gtrsim Q$.  This preserves the required approximation property of
$\TTnew{}{}$ at small $\Tsc{q}$.  At the same time, since the changes
are dominantly at large $\Tsc{q}$, the integral over all $\T{q}$ still
has a normal collinear expansion, as we will make more explicit below.

A simple $\Theta(Q - \Tsc{q})$ step function is acceptable for
$\Xi$. When we combine $\TTnew{}{}$ with a $Y$-term in
Secs.~\ref{sec:Yterm}--\ref{sec:together} we will introduce methods to
minimize sensitivity to the exact form of
$\Xi\xleft(\Tsc{q}/Q,\eta\right)$.  However, a smoother function is
preferred since the domain of validity of the $W$-term approximation
does not end at a sharp point in $\Tsc{q}$, and thus a smooth function
characterizes general physical expectations.  A reasonable choice is
\begin{equation}
\Xi\xleft(\frac{\Tsc{q}}{Q},\eta\right) = \exp \left[ -\left( \frac{q_T}{ \eta Q} \right)^{a_\Xi}   \right] \, , \label{eq:Xiparam}
\end{equation}
with $a_\Xi > 2$.  

The only differences between the old and new $W$-term are: i) the use
of $\bone(\Tsc{b})$ rather than $\Tsc{b}$ in $\tilde{W}$, and ii) the
multiplication by $\Xi\xleft(\Tsc{q}/Q,\eta\right)$. (The second
modification was proposed by Collins in
Ref.~\cite[Eq.~(13.75)]{Collins:2011qcdbook}. There $\Xi$ is called
$F(\Tsc{q}/Q)$.) Equation~\eqref{eq:Wnew} matches the standard
definition in the limit that $C_5$ and $\eta$ approach infinity.

Finally, we will present a fully optimized formula for $\TTnew{}{}$ corresponding
to the one for the original $\TT{}{}$ in Eq.\ (\ref{eq:finalevolved}).

But first it will be convenient to construct some auxiliary results. 

Naturally, $\bstarsc$ is to be replaced by
\begin{equation}
  \bstarsc(\bone(\Tsc{b}))
  = \sqrt{ \frac{ \Tsc{b}^2 + b_0^2/(C_5^2Q^2) }
            { 1 + \Tsc{b}^2/\bmax^2 + b_0^2/(C_5^2Q^2\bmax^2) }
         } \, .
\end{equation}
Also we define
\begin{equation}
\bmin \equiv \bstarsc(\bone(0))
= \frac{b_0}{C_5Q} \sqrt{\frac{1}{1 + b_0^2/(C_5^2Q^2\bmax^2)}} \, .
\end{equation}
Then, for large enough $Q$ and $\bmax$
\begin{equation}
\bmin \approx \frac{b_0}{C_5Q} \, .
\end{equation}
Thus, $\bmin$ decreases like $1/Q$, in contrast to $\bmax$ which remains fixed.
Note also that
\begin{equation}
\bstarsc(\bone(\Tsc{b})) \longrightarrow
\begin{dcases}
\bmin & \Tsc{b} \ll \bmin \\
\Tsc{b} & \bmin \ll \Tsc{b} \ll \bmax \\
\bmax & \Tsc{b} \gg \bmax \, . \label{eq:bdef2}
\end{dcases}
\end{equation}
For $\Tsc{b} \ll 1/Q$,
$\bstarsc(\bone(\Tsc{b})) \approx \bstarsc(\Tsc{b})$.
Instead of $\mubstar$, we will ultimately use the scale 
\begin{equation}
\mucirc \equiv \frac{C_1}{\bstarsc(\bone(\Tsc{b}))} \, 
\end{equation}
to implement renormalization group improvement in TMD correlation functions.
There is a maximum cutoff on the renormalization scale equal to 
\begin{equation}
\mucircm
\equiv \lim_{\Tsc{b} \to 0} \mucirc
= \frac{C_1C_5 Q}{b_0} \sqrt{1 + \frac{b_0^2}{C_5^2\bmax^2 Q^2}}
\approx \frac{C_1C_5Q}{b_0}
\, . \label{eq:mucut}
\end{equation}
The approximation sign corresponds to the limit of large $Q\bmax$.
Note that,
\begin{equation}
\bmin \mucircm = C_1 \, .
\end{equation}

\begin{widetext}

The steps for finding a useful formula for the evolved
$\TTnew{}{}$ are as follows.
Equation~\eqref{eq:solngb} becomes
\begin{equation}
\TTnew{}{} = \Xi\xleft(\frac{\Tsc{q}}{Q},\eta\right) \int \frac{\diff[2]{\T{b}}}{(2 \pi)^2}
    e^{i\T{q}\cdot \T{b} } \tilde{W}_{\rm NP}(\bone(\Tsc{b}),Q) \tilde{W}(\bstarsc(\bone(\Tsc{b})),Q) \, . \label{eq:evolWL}
\end{equation}
Now the definition of $\tilde{W}(\Tsc{b},Q)$ is unchanged,
and only the $\Tsc{b} \to \bone(\Tsc{b})$ replacement is new.  
Therefore instead of 
Eq.~\eqref{eq:finalevolved} we simply need
\begin{eqnarray}
\tilde{W}(\bone(\Tsc{b}),Q)  
 & = & H(\mu_{Q},Q) \sum_{j' i'} \int_{x_A}^1
\frac{d \hat{x}}{\hat{x}}  \tilde{C}^{\rm pdf}_{j/{j'}}(x_A/\hat{x},\bstarsc(\bone(\Tsc{b}));\mucirc^2,\mucirc,\alpha_s(\mucirc)) f_{j'/A}(\hat{x};\mucirc) \times \nonumber \\
& & \times  \int_{z_B}^1 \frac{d \hat{z}}{\hat{z}^3} \tilde{C}^{\rm ff}_{i'/{j}}(z_B/\hat{z},\bstarsc(\bone(\Tsc{b}));\mucirc^2,\mucirc,\alpha_s(\mucirc)) d_{B/i'}(\hat{z};\mucirc) \times \nonumber \\
& \times & \exp \left\{ \ln \frac{Q^2}{\mucirc^2} \tilde{K}(\bstarsc(\bone(\Tsc{b}));\mucirc) + 
\int_{\mucirc}^{\mu_{Q}} \frac{d \mu^\prime}{\mu^\prime} \left[ 2 \gamma(\alpha_s(\mu^\prime);1) 
- \ln \frac{Q^2 }{{\mu^\prime}^2} \gamma_K(\alpha_s(\mu^\prime)) \right]\right\}  \nonumber \\ 
 & \times &  \exp\left\{ -g_{A}(x_A,\bone(\Tsc{b});\bmax)-g_{B}(z_B,\bone(\Tsc{b});\bmax)  
-2 g_K(\bone(\Tsc{b});\bmax) \ln \left( \frac{Q}{Q_0} \right) \right\} \, . \label{eq:finalevolvedb}
\end{eqnarray}
This is the same as Eq.~\eqref{eq:finalevolved} except that $\bstarsc(\bone(\Tsc{b}))$ and $\mucirc = C_1 / \bstarsc(\bone(\Tsc{b}))$ are 
used instead of $\bstarsc(\Tsc{b})$ and $\mubstar = C_1 / \bstarsc(\Tsc{b})$.  
Note that $g_K(\bone(\Tsc{b});\bmax)$ depends
on $Q$ through $\bone$, albeit only for $\Tsc{b} \lesssim 1/Q$.
For $\Tsc{b} \gg 1/Q$, $g_K(\bone(\Tsc{b});\bmax) \to g_K(\Tsc{b};\bmax)$. 
Also,  $g_K(\bone(\Tsc{b});\bmax)$ does not 
vanish exactly as $\Tsc{b} \to 0$ but instead approaches a power of $1/Q$.

Up to this point, we have introduced two new parameters, $\eta$ and
$C_5$, in the treatment of the $W$-term. 

\section{Modified $Y$-Term}
\label{sec:Yterm}

Now we can construct a $Y$-term from nearly identical steps to those of Sec.~\ref{sec:largesmall}. 
Recall that the TMD approximator, $\appor{TMD}$, 
replaces the cross section by an approximation that is good in the
$\Tsc{q} / Q \ll 1$ limit -- see Eq.~\eqref{eq:wterm}.
The $\appor{TMD}$ from Ref.~\cite{Collins:2011qcdbook} replaces $\cs{}{}$ by the definition of 
$W(\Tsc{q},Q)$ that follows most directly from the derivation of TMD
factorization. However, any approximator that is good when
  $\Tsc{q} \ll Q$ is equally valid here.
Therefore, we write
\begin{equation}
\label{eq:wtermmod2}
\TTnew{}{} \equiv \appor{TMD}^{\rm New} \cs{}{}  \,. 
\end{equation}
Here $\appor{TMD}^{\rm New}$ applies the same approximations as
$\appor{TMD}$, but with the use of
  $\Xi\xleft(\Tsc{q}/Q,\eta\right) $ and $\bstarsc(\bone(\Tsc{b}))$ as
in Eq.\ (\ref{eq:Wnew}).

Since the changes only affect the region
$\Tsc{q} \gtrsim Q$, 
power counting for small $\Tsc{q}$ proceeds in exactly the same way as in
Sec.~\ref{sec:largesmall}:
\begin{equation}
\appor{TMD}^{\rm New} \cs{}{} = \cs{}{} 
                + O\xleft( \frac{\Tsc{q}}{Q}\right)^a \cs{}{}
                + O\xleft( \frac{m}{Q} \right)^{a'} \cs{}{}
 \,.
\label{eq:TMDapdef2}
\end{equation} 
The large $\Tsc{q} \sim Q$ region is dealt with using the same $\appor{coll}$ approximator as in Sec.~\ref{sec:largesmall}: 
\begin{equation}
\appor{coll} \cs{}{} = \cs{}{} 
    + O\xleft( \frac{m}{\Tsc{q}} \right)^b \cs{}{}
\, .
\label{eq:collapdef2}
\end{equation} 
Continuing the usual steps, a $Y$-term is constructed by adding and subtracting $\TTnew{}{}$:
\begin{equation}
  \cs{}{} \, =  \appor{TMD}^{\rm New} \cs{}{}  \
       + \Bigg[ \cs{}{} - \appor{TMD}^{\rm New} \cs{}{} \Bigg]
\, .
\label{eq:nextapp2}
\end{equation}
The term in brackets is
only unsuppressed for large $\Tsc{q}$, so we apply to it the 
large $\Tsc{q}$ approximator, $\appor{coll}$,  
and use collinear factorization: 
\begin{align}  
\Gamma(m \lesssim \Tsc{q},Q)  ={}&
   \appor{TMD}^{\rm New} \cs{}{} \
   + \appor{coll}
    \left[ \cs{}{} - \appor{TMD}^{\rm New} \cs{}{} \right]
\no &
 + O\xleft( \left( \frac{m}{\Tsc{q}} \right)^b \left( \frac{\Tsc{q}}{Q} \right)^a  \right) \cs{}{} 
  + O\xleft( \left( \frac{m}{\Tsc{q}} \right)^b \left( \frac{m}{Q} \right)^{a'}  \right) \cs{}{} 
\nonumber \\
  ={}& \TTnew{}{} 
       + \appor{coll}\cs{}{} -  \appor{coll}\appor{TMD}^{\rm New}
         \cs{}{}
+ O\xleft(  \frac{m}{Q}\right)^{\rm c} \cs{}{}
\, ,
\label{eq:powercounting2}
\end{align}
where $c = {{\rm min}(a,a',b)}$.

Finally, we insert a factor of the $X(\Tsc{q}/\lambda)$ function from Eq.~\eqref{eq:yterm} to remove any $Y$-term
contribution in the $\Tsc{q} < m$ region. The final $Y$-term is
\begin{align}
\label{eq:yterm2}
\YYnew{}{} 
  &{}\equiv \left\{ \appor{coll} \left[ \cs{}{} - \TTnew{}{} \right]
            \right\} X(\Tsc{q}/\lambda)
 \nonumber \\
   &{}= \left\{ \appor{coll} \cs{}{} - \appor{coll} \appor{TMD}^{\rm New} \cs{}{} \right\} X(\Tsc{q}/\lambda) \, . 
\end{align}
Then,
\begin{align}
\fixo{}{} & \equiv \appor{coll} \cs{}{} \label{eq:fixocal} \\
\asnew{}{} &\equiv \appor{coll} \appor{TMD}^{\rm New} \cs{}{}  \label{eq:asydef2} \, .
\end{align}
So,
\begin{equation}
\YYnew{}{} = \left\{ \fixo{}{} - \asnew{}{}  \right\} X(\Tsc{q}/\lambda) \, . \label{eq:finalydef2}
\end{equation}
As usual, $\appor{coll}$ is an instruction to set all renormalization scales to $\mu = \mu_Q$ and drop 
powers of $m/\Tsc{q}$ or $m/Q$. 
In $\appor{coll} \appor{TMD}^{\rm New} \cs{}{}$, the $\appor{TMD}^{\rm New}$ inserts a multiplication by 
a factor of $\Xi(\Tsc{q}/Q,\eta)$, effectively setting $\appor{TMD}^{\rm New} \cs{}{}$ to zero for 
large $\Tsc{q} \gtrsim Q$ (see, e.g., Eq.~\eqref{eq:Xiparam}).  Thus, if $\Xi(\Tsc{q}/Q,\eta)$ gets dropped when 
$\appor{coll}$ is applied, there is a potential to introduce large errors. Therefore, $\appor{coll}$ should \emph{not} drop the 
factor of $\Xi(\Tsc{q}/Q,\eta)$ because $\appor{coll}$, by definition, must leave the $\Tsc{q} \gg m$ region unmodified. Similarly, 
the use of $\bone(\Tsc{b})$ affects the small $\Tsc{b}$ limit of $\bstarsc(\bone(\Tsc{b}))$, and therefore can also have a large effect on 
on $\appor{TMD}^{\rm New} \cs{}{}$ at large $\Tsc{q}$. Thus,
$\appor{coll}$ should preserve the use of $\bone(\Tsc{b})$. 
In contrast, $\bmax \sim 1/m$ mainly affects the small $\Tsc{q}$
region. Therefore,
we define $\appor{coll}$ to apply the 
$\bmax \to \infty$ limit in Eq.~\eqref{eq:asydef2}. Examples of implementations of Eqs.~\eqref{eq:fixocal}--\eqref{eq:finalydef2} will be 
given in Secs.~\ref{sec:bcfg} and~\ref{sec:demo}.  

  Now observe that $\Xi(\Tsc{q}/Q,\eta)$ approaches zero as $\Tsc{q}$ gets
  much larger than $Q$.  Then $\YYnew{}{}$ approaches the usual collinear
  factorization result for $\cs{}{}$ at large $\Tsc{q}$.
Therefore, we may at last remove the $\Tsc{q} \lesssim Q$ restriction on the 
left side of Eq.~\eqref{eq:basic2} and write a $W+Y$ representation of the cross section that extends over the whole range of $\Tsc{q}$:
\begin{equation}
\label{eq:basic22}
\Gamma(\Tsc{q},Q)  = \TTnew{}{} + \YYnew{}{} + O\xleft( \frac{m}{Q}\right)^{\rm c} \cs{}{} \, .
\end{equation} 
We have reached our goal of constructing a $W+Y$ expression that does not require that we  
specify limitations on the range of $\Tsc{q}$. What remains is to 
determine the most appropriate values for $\eta$ and $C_5$.

\section{Connection with $\T{q}$-integrated cross sections and collinear factorization}
\label{sec:together}

In this section, we analyze the integral over all $\T{q}$ of the
right-hand side of Eq.\ (\ref{eq:basic22}), and show how it matches
standard collinear factorization for the integrated cross section.

 We integrate Eq.~\eqref{eq:basic22} over all transverse momentum,
  and then reorganize the result as follows:
\begin{align}
\int \diff[2]{\T{q}} \cs{}{} ={}& \int \diff[2]{\T{q}} \left[ \TTnew{}{} +  \YYnew{}{}  \right] \, \nonumber \\
	={}& \int \diff[2]{\T{q}} \left[ \Xi\xleft(\frac{\Tsc{q}}{Q},\eta\right) 
  \int \frac{\diff[2]{\T{b}}}{(2 \pi)^2}
    e^{i\T{q}\cdot \T{b} } \tilde{W}(\bone(\Tsc{b}),Q) +  \YYnew{}{}  \right]  \, \nonumber \\ \nonumber \\ 
   ={}&  \int \diff[2]{\T{q}} \int \frac{\diff[2]{\T{b}}}{(2 \pi)^2}
    e^{i\T{q}\cdot \T{b} } \tilde{W}(\bone(\Tsc{b}),Q) &  \qquad \qquad {\rm Term \, 1} \nonumber \\
   {}&-  \int \diff[2]{\T{q}} \left( 1 - \Xi\xleft(\frac{\Tsc{q}}{Q},\eta\right) \right) \int \frac{\diff[2]{\T{b}}}{(2 \pi)^2}
    e^{i\T{q}\cdot \T{b} } \tilde{W}(\bone(\Tsc{b}),Q)  &  \qquad \qquad {\rm Term \, 2} \nonumber \\
   {}&+  \int \diff[2]{\T{q}} \YYnew{}{} \, . &  \qquad \qquad {\rm Term \, 3}  \nonumber \\
\end{align}
  Term 1 is $\TTnew{}{}$ integrated over $\T{q}$, but without the $\Xi$ factor, so it can easily
  be simplified.  Term 2 corrects for the omission of $\Xi$, while
  term 3 is the integral of the $Y$ term.

 Now term 1 equals $\tilde{W}(\bone(0),Q) = \tilde{W}(\bmin,Q)$.  Since
$\bmin=O(1/Q)$, we can replace it by the OPE $\tilde{W}^{\rm
  OPE}(\bmin,Q))$ form, to leading power in $m/Q$, Eq.\
(\ref{eq:smallblimit}), to obtain
\begin{equation}
  \mbox{Term 1}
  = \tilde{W}^{\rm OPE}(\bmin,Q) 
    + O\xleft( ( m/Q )^p \right) .
\end{equation}
Then, we can use Eq.\ (\ref{eq:Tcoll}) to give a factorization in
terms of collinear pdfs and ffs at a scale of order $Q$.  Since in
that formula $\bstarsc(\Tsc{b})$ is replaced by $\bmin=O(1/Q)$, while 
$\mubstar$ is of order $Q$, both the $\tilde{C}$ factors and the
quantities in the exponential can be expanded in powers of
$\alpha_s(Q)$ without large logarithms. We therefore have a normal
collinear expansion.  The lowest-order term gives
\begin{equation}
  \mbox{Term 1}
  = H_{{\rm LO}, \, j j} f_{j/A}(x;\mucircm) \, d_{B/j}(z;\mucircm)
    + O(\alpha_s(Q)),
\end{equation}
with our choice of scale given in Eq.\ (\ref{eq:mucut}).  This agrees
with the lowest-order term for the integrated cross section itself,
i.e., for $\int \diff[2]{\T{q}} \cs{}{}$. 

Both terms 2 and 3 are dominated in their integrals by $\Tsc{q}$ of
order $Q$.  They therefore have normal collinear expansions, starting
at order $\alpha_s(Q)$.
Overall, we therefore have well-behaved perturbative expansions of
collinear factorization for each term, unlike the case for the
$\Tsc{q}$ integrals of the original CSS forms for $W$ and $Y$.

We now show more explicitly that terms 2 and 3 are dominated by
$\Tsc{q}$ of order $Q$.  For term 2, the factor $1-\Xi$ gives a power
suppression for $\Tsc{q} \ll Q$, while the use of $\bone(\Tsc{b})$
instead of $\Tsc{b}$ gives an exponential suppression for $\Tsc{q} \gg
Q$, as we have already seen.  For term 3, the construction of
$\YYnew{}{}$ gives power suppression when $\Tsc{q} \ll Q$, with the
factor $X(\Tsc{q}/\lambda)$ in (\ref{eq:finalydef2}) ensuring that no
pathologies arise when $\Tsc{q}$ is very small (below $m$).  At large
$\Tsc{q}$, beyond $Q$, the $\fixo{}{}$ term obeys the kinematic limit, while
the $\asnew{}{}$ term is exponentially suppressed, for the same reason
as for $\TTnew{}{}$. 

\end{widetext}

\section{Calculating the asymptotic term in the BCFG method}
\label{sec:bcfg}

  Perturbative calculations for the hard coefficient for the $Y$
  term in the original CSS version can be performed by starting from
  the normal collinear coefficient for the cross section as a function
  of $\Tsc{q}$.  Then the asymptote at small $\Tsc{q}$ is subtracted.
  This asymptote is simply the leading power expansion in $\Tsc{q}/Q$ when $\Tsc{q}$ is
  much smaller than $Q$, and involves simply a factor of $1/\Tsc{q}^2$
  time logarithms of $Q/\Tsc{q}$ in each order of perturbation theory.
  The coordinate-space version of the subtraction in each order is
  correspondingly a polynomial in $\ln (Q\Tsc{b})$.

  In the new scheme, the coordinate space formula is unchanged, but it
  is not so simple to perform a practical analytic calculation of its
  Fourier transform to give $\asnew{}{}$.  This is because of the
  substitution of $\bone(\Tsc{b})$ for $\Tsc{b}$.  We now explain how
  to do this, following Ref.~\cite{Bozzi:2005wk}.

  Calculations of $\asnew{}{}$ need Fourier-Bessel transforms of
  terms of the form
\begin{multline}
\label{eq:logs2}
\alpha_s(\muQ)^m \ln^n \xleft( \frac{\muQ^2 \bone(\Tsc{b})^2}{b_0^2} \right) =
\\
\alpha_s(\mu_Q)^m \ln^n \xleft( \frac{\muQ^2 \Tsc{b}^2}{b_0^2} + \frac{C_2^2}{C_5^2} \right) \, .
\end{multline}
with $m \geq 1$ and $0 \leq n \leq 2m$ and $b_0 \equiv 2
\exp(-\gamma_E)$.  (The use of $b_0$ in the argument of the
  logarithm is a convention that typically results in simpler
  formulas.)
  These terms arise from the perturbative expansion $\appor{coll}
  \appor{TMD}^{\rm New} \cs{}{}$.  This can be considered as arising
  from the collinear factorization of Eq.\ (\ref{eq:Tcoll}) with
  $\mubstar$ replaced by $\muQ$, with all couplings expressed in terms
  of $\alpha_s(\muQ)$, and then with a fixed-order perturbative
  expansion applied to the product of the $\tilde{C}$ factors and the
  exponential in Eq.\ (\ref{eq:Tcoll}).

If $Q^2 \Tsc{b}^2 \gg C_2^2/ C_5^2$, then we neglect the second term in the logarithms, and Eq.~\eqref{eq:logs2} becomes 
the much more familiar form from standard CSS-like treatments 
\begin{equation}
\label{eq:logsreduce}
\alpha_s(\mu_Q)^m \ln^n \xleft( \frac{\mu_Q^2 \Tsc{b}^2}{b_0^2}  \right) \, .
 \end{equation}

\subsection{Standard Logarithms} 

In the CSS and related treatments, with the standard $W+Y$ construction, the logarithms are of the form of Eq.~\eqref{eq:logsreduce}.
In that case, the momentum space expressions are well-known (see, e.g., Eq.~(36) of Ref.~\cite{Nadolsky:1999kb}). 
After Fourier transformation, coordinate space logarithmic terms
like Eq.~\eqref{eq:logsreduce} give $\Tsc{q}$-dependence like
\begin{equation}
\frac{1}{\Tsc{q}^2} \, , \, \frac{1}{\Tsc{q}^2} \ln \xleft( \frac{Q^2}{\Tsc{q}^2} \right) \, , \, \dots \label{eq:momlogs}
\end{equation}
where the ``$\dots$'' refers to higher power logarithms. 

\subsection{Modified logarithms}

A primary motivation for our modified $W+Y$ construction is to
accommodate a non-zero $\bmin$ in
Eqs.~\eqref{eq:evolWL} and \eqref{eq:finalevolvedb},
and thus a non-zero $C_2/C_5$ in Eq.~\eqref{eq:logs2}.   
Fortunately, for the case of non-zero $C_2/C_5$, analytic expressions for 
the finite parts of the Fourier-Bessel transforms have been worked out in Appendix B of BCFG, Ref.~\cite{Bozzi:2005wk}. 
Indeed, the case of of $C_5 = C_2$ corresponds exactly to the 
$\ln^m (Q^2 \Tsc{b}^2/b_0^2) \to \ln^m (Q^2 \Tsc{b}^2/b_0^2 + 1)$
prescription of PP~\cite{Parisi:1979se} and used in implementations
like~\cite{Bozzi:2005wk}. 

  Now, the discussion so far has been based on the expression for
  $\TTnew{}{}$ in terms of TMD densities.  However, to get the hard
  coefficient for $\asnew{}{}$, as needed in $\YYnew{}{}$, it is also
  possible to start from the hard coefficients for ordinary collinear
  factorization for the cross section.  Then one does the expansion at
  small $\Tsc{q}$ to give $1/\Tsc{q}^2$ times logarithms.  Finally to
  obtain the effect of the use of $\bone(\Tsc{b})$ instead of
  $\Tsc{b}$, one makes the substitutions given below.

One can read the substitutions off from 
results like Ref.~\cite[Eqs.~(B.10)-(B.13)]{Bozzi:2005wk}. For example,
\begin{align}
 \frac{1}{\Tsc{q}^2} 
 \to{}& \frac{C_2 b_0}{\Tsc{q} \mu_Q C_5} K_1 \xleft( \frac{C_2 \Tsc{q} b_0}{C_5 \mu_Q} \right) \,
\label{eq:logrep1} \\
 \frac{1}{\Tsc{q}^2} \ln \xleft(  \frac{\mu_Q^2}{\Tsc{q}^2} \right) 
 \to{}&
 \frac{C_2 b_0}{\Tsc{q} \mu_Q C_5} 
 \left[ K_1 \xleft( \frac{C_2 \Tsc{q} b_0}{C_5 \mu_Q} \right) \ln \xleft( \frac{C_2 \mu_Q}{C_5
       \Tsc{q}} \right) + \right.
 \nonumber \\ 
 & \qquad \qquad
 + \left.  K_1^{(1)}\xleft( \frac{C_2 \Tsc{q} b_0}{C_5 \mu_Q} \right) \right] \, .
\label{eq:logrep2}
\end{align}
Here, $K_\nu(x)$ is the modified Bessel function of the second kind and
\begin{equation}
K_1^{(1)}(x) \equiv \left. \frac{\partial}{\partial \nu} K_\nu(x) \right|_{\nu = 1} \, .
\end{equation}
The left and right sides of Eqs.~\eqref{eq:logrep1}--\eqref{eq:logrep2} are approximately equal for fixed $C_5$ and $\Tsc{q} \ll \mu_Q$.
See also the discussion around Ref.~\cite[Eq.~(B.25)]{Bozzi:2005wk}.

Now one may perform substitutions like Eqs.~\eqref{eq:logrep1}--\eqref{eq:logrep2} to
known results for the asymptotic term like Eq.~(36) of Ref.~\cite{Nadolsky:1999kb} to obtain the generalized, non-zero $C_2/C_5$, asymptotic term. 
Reference~\cite[Appendix B]{Bozzi:2005wk} contains results for any $n$, so the modified asymptotic term, and 
thus the new $Y$-term, can be obtained to any order from previously existing expressions. 
For completeness, low order expressions for the asymptotic terms are given in Appendix~\ref{sec:asy}.

\section{Demonstration}
\label{sec:demo}

To illustrate the steps above, we have performed sample calculations of the $Y$-term using 
analytic approximations for the collinear pdfs and collinear ffs. For simplicity, we consider only the target up-quark $\gamma^\ast q \to q g$ channel, 
and for the running $\alpha_s(\mu)$ we use the two-loop $\beta$-function solution and keep the number of flavors at $n_f  = 3$ since
we are mainly interested in the transition to low $Q$. Thus we use $\Lambda_{\rm QCD} = 0.339$~GeV~\cite{Bethke:2012jm}. 
To further simplify our calculations, we use analytic expressions for the collinear correlation functions, taken from appendix A1 of Ref.~\cite{Gluck:1991ng} for the 
up-quark pdf and from Eq.~(A4) of Ref.~\cite{Kniehl:2000hk} for the up-quark-to-pion fragmentation function.

Due to these simplifying assumptions, the following should be regarded as a 
toy model calculation, meant to 
illustrate the basic steps of a $Y$-term calculation and to demonstrate plausibility for use in more complete and detailed calculations.
\begin{figure}
\centering
  \begin{tabular}{c@{\hspace*{10mm}}c}
    \includegraphics[scale=0.3]{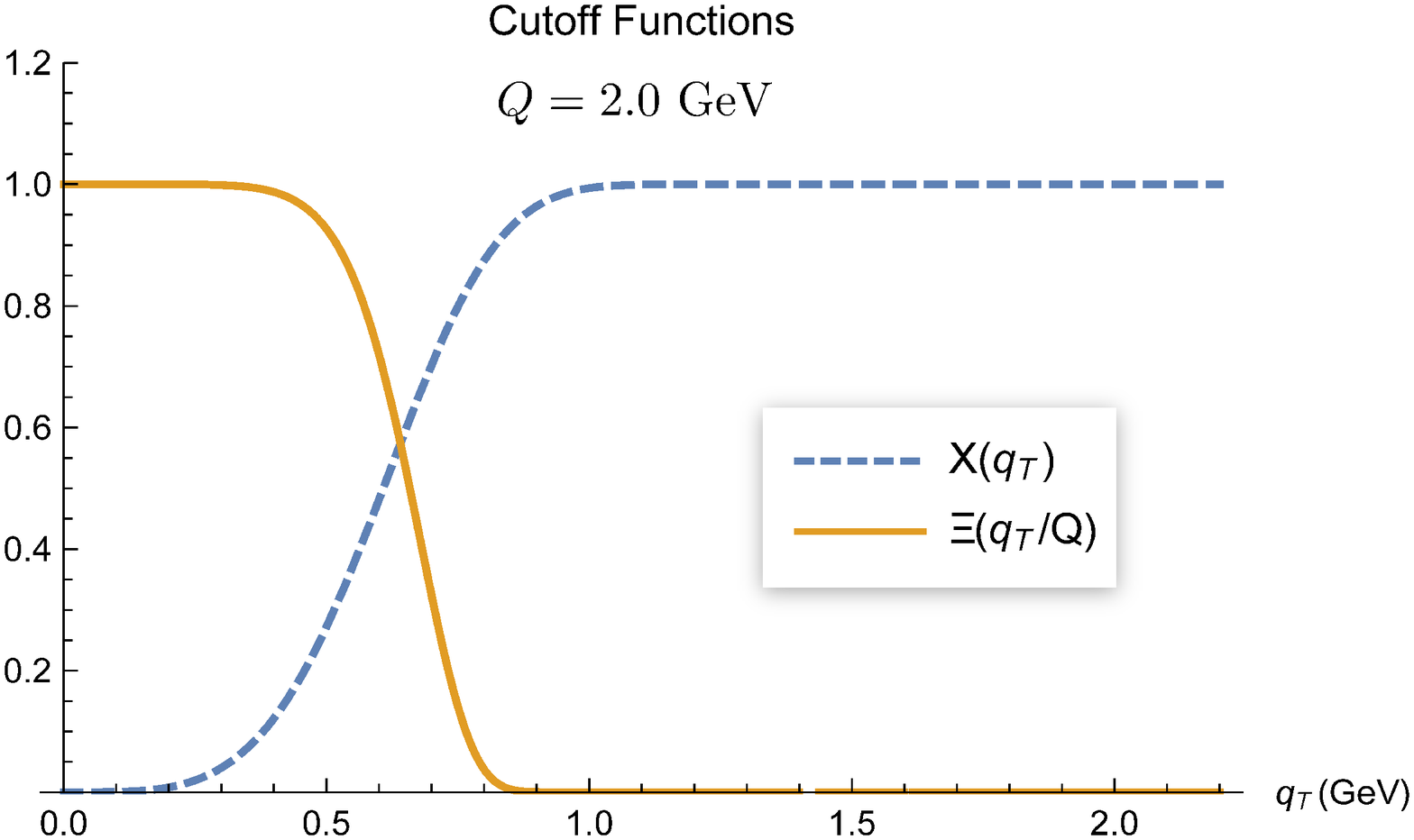} \\
    (a) \\
    \vspace{8mm} \\
    \includegraphics[scale=0.3]{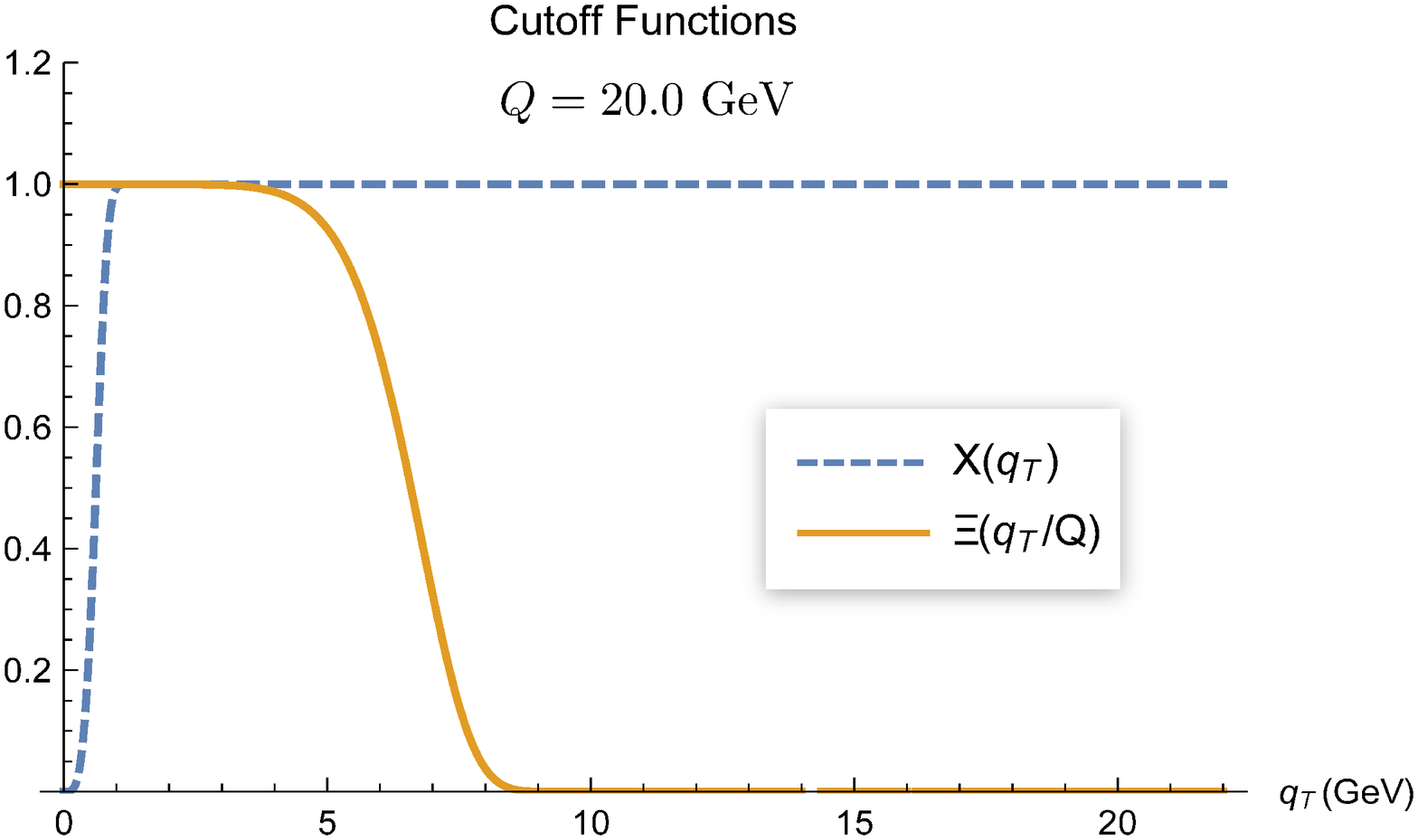}
    \\
    (b)
    \\[5mm]
   \end{tabular}
\caption{The cutoff functions in Eq.~\eqref{eq:Xparam} for low $\Tsc{q} / \lambda$ (blue dashed line) and 
in Eq.~\eqref{eq:Xiparam} for large $\Tsc{q}/Q$ (brown solid line) for $Q = 20.0$~GeV (plot (a)) and $Q = 2.0$~GeV (plot (b)). 
In both, $\lambda = 2/3$~GeV and $\eta = 0.34$. The region of $\Tsc{q} \gtrsim Q/4$ is determined by the $\fixo{}{}$ calculation. 
For all $Q$, $\Tsc{q} \lesssim \lambda$ is considered non-perturbative. (Color online.) 
}
\label{fig:cutoffs}
\end{figure}

First, one must establish parameters for our large and small $\Tsc{q}$ cutoff 
functions. For $X(\Tsc{q}/\lambda)$ we use Eq.~\eqref{eq:Xparam}, and try $a_X = 4$ since this gives a rapid but reasonably gentle suppression of small $\Tsc{q}$. The choice 
of $\lambda$ should be such that it has reached unity at values of $\Tsc{q}$ near the 
perturbative-nonperturbative transition, say, $\Tsc{q} \approx 1.0$~GeV. Thus, we choose $\lambda = 2/3$~GeV.
The result is shown as the blue dashed curves in Figs.~\ref{fig:cutoffs}. To understand the plots, recall that $X(\Tsc{q}/\lambda)$ is used to 
restrict to large $\Tsc{q}$ the region where $\Tsc{q}$-dependence is calculated with collinear factorization at fixed order fixed in perturbation theory.

For $\Xi\xleft(\Tsc{q}/Q,\eta\right)$ we use Eq.~\eqref{eq:Xiparam}. The value of $a_\Xi$ controls how rapidly 
the $\Tsc{q} \sim Q$ contribution from the $W$-term gets cutoff. For large $Q$, the transition can be rather 
smooth since there is a broad region where $\asnew{}{}$ and $\fixo{}{}$ overlap. In our example calculation, we find that $a_\Xi = 8$ works well.
The value of $\eta$ should be chosen such that  $\Xi\xleft(\Tsc{q}/Q,\eta\right) \to 0$ when $\Tsc{q}$ is large enough that 
approximations that use $\Tsc{q} \ll Q$ might be considered suspect. For small $\Tsc{q}$, 
$\Xi\xleft(\Tsc{q}/Q,\eta\right) \to 1$. We find that the transition between $\Xi\xleft(\Tsc{q}/Q,\eta\right) \approx 0$ 
and $\Xi\xleft(\Tsc{q}/Q,\eta\right) \approx 1$ occurs between about $\Tsc{q} \approx Q/4$ and $\Tsc{q} \approx Q/2$ if $\eta = 0.34$.
These results for $\Xi\xleft(\Tsc{q}/Q,\eta\right)$ are shown as the tan curves in Figs.~\ref{fig:cutoffs}. 
To understand the plots, recall that the purpose of $\Xi\xleft(\Tsc{q}/Q,\eta\right)$ is to 
suppress the $\Tsc{q} = O(Q)$ region of the $W$-term where it fails to provide even a rough approximation.

\begin{figure}
\centering
    \includegraphics[scale=0.3]{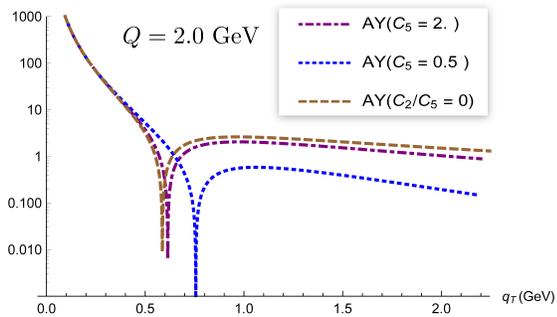}
\caption{The absolute value of the asymptotic term calculation with $\Xi$ replaced by 1, and with the substitutions in Eq.~\eqref{eq:logrep1}--\eqref{eq:logrep2} and various choices for $C_5$. The brown dashed curve 
is the limit of the standard CSS $Y$-term approach. In all cases, $C_2 = 1$. The blue dotted and magenta dash-dotted curves correspond $C_5= 0.5$ and $C_5 = 2.0$ respectively. All curves 
are normalized to ${\rm FO}(q_{\rm T},Q)$ and $\Tsc{q} = 1$~GeV. The variation 
between the curves can be viewed as an measure of the sensitivity of the $\as{}{}$ calculation to different choices of $C_5$. (Color online.) In all cases, we take $x = 0.1$ and $z = 0.5$.
}
\label{fig:asymplots}
\end{figure}
\begin{figure}
\centering
  \begin{tabular}{c@{\hspace*{10mm}}c}
    \includegraphics[scale=0.3]{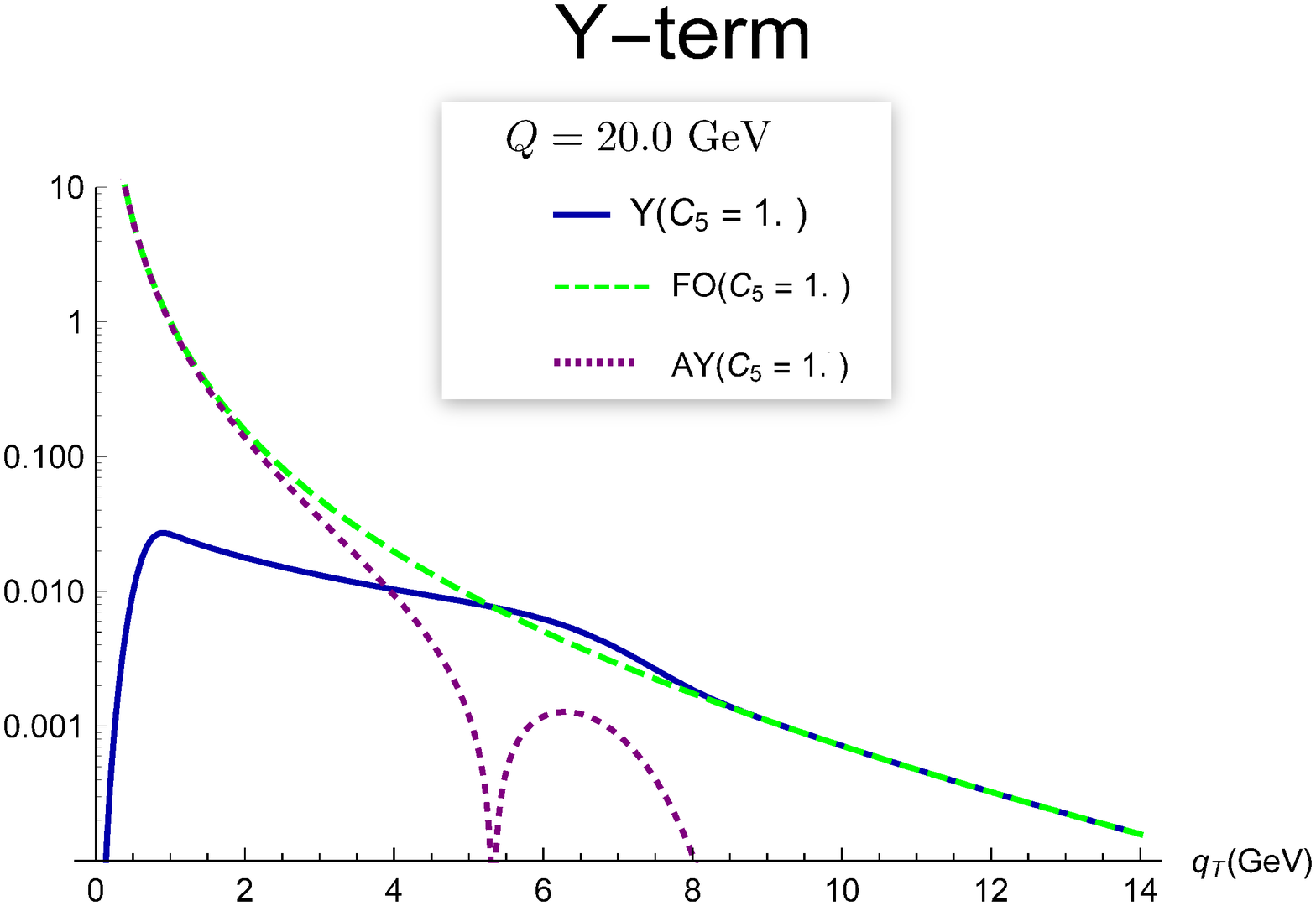} \\
    (a) \\
    \vspace{8mm} \\
    \includegraphics[scale=0.31]{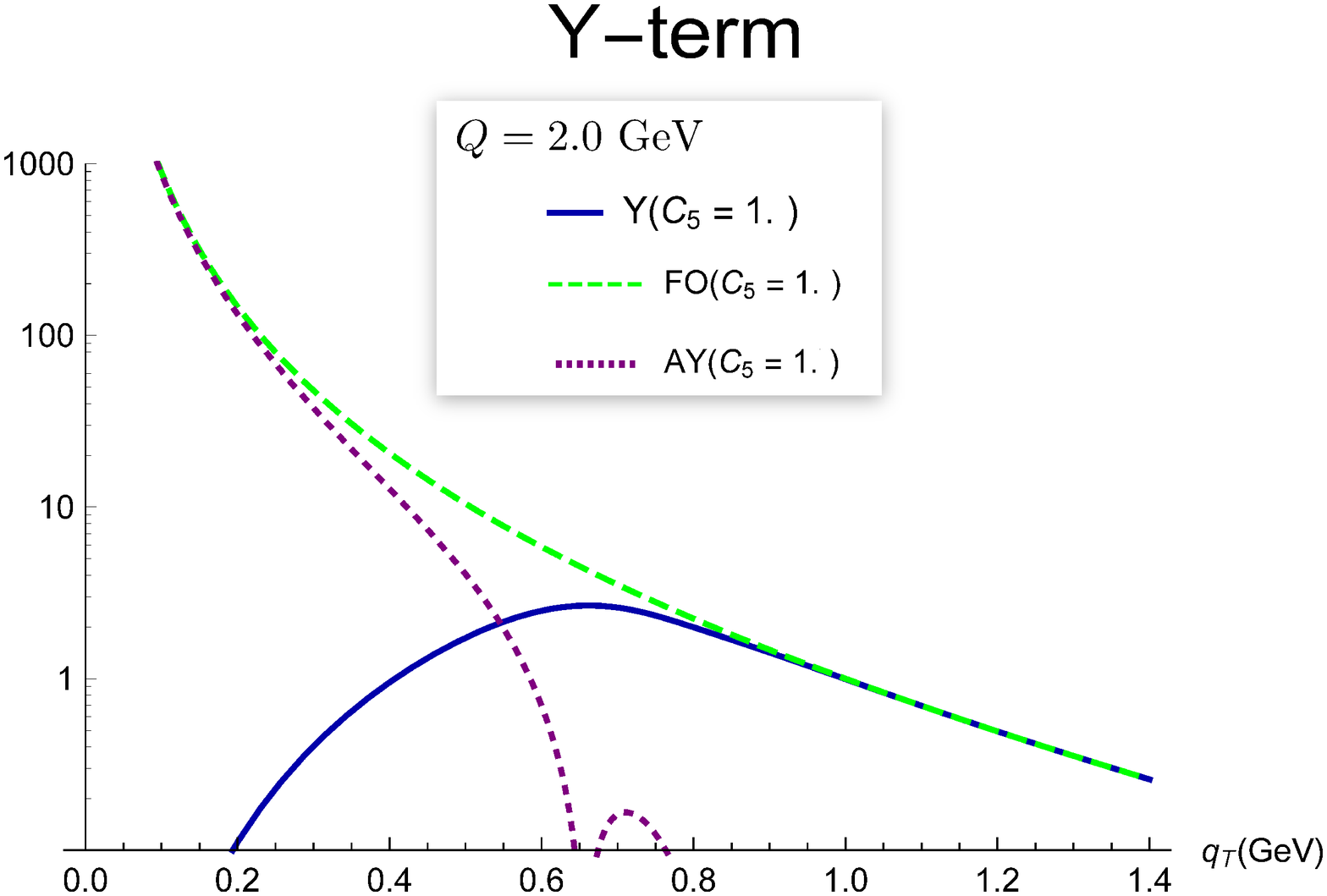}
    \\
    (b)
    \\[5mm]
   \end{tabular}
\caption{The Y-term (blue solid curves) calculated using the method of Eq.~\eqref{eq:finalydef2} and 
Sec.~\ref{sec:bcfg}. One calculation (a) is for a large scale, $Q=20.0$~GeV and one calculation (b) is for a small 
scale, $Q = 2.0$~GeV. For comparison, the $\fixo{}{}$ (green dashed) and $\asnew{}{}$ (magenta dot-dashed) calculations are also shown. In all cases, $C_5 = 1.0$. The curves 
are normalized to the value of ${\rm FO}(q_{\rm T},Q)$ at $\Tsc{q} = 1.0$~GeV. (Color online.) In all cases, we take $x = 0.1$ and $z = 0.5$.
}
\label{fig:yterms}
\end{figure}
Next, we examine the effect of varying $C_5$ on the calculation of the asymptotic term. 
Standard expressions for the asymptotic term can be found in, for example, 
Eq.~(36) of Ref.~\cite{Nadolsky:1999kb}. We use these results, along with the substitutions in 
Eqs.~\eqref{eq:logrep1}--\eqref{eq:logrep2}, to plot the new asymptotic term of Eq.~\eqref{eq:asydef2} for a range of $C_5$ values. 
The result is shown in Fig.~\ref{fig:asymplots}, where 
we have temporarily set $\Xi\xleft(\Tsc{q}/Q,\eta\right)$ to $1$
in order to highlight the effect of varying $C_5$. The results for $C_5 = 0.5$ and $C_5 = 2.0$ are shown. 
The standard CSS result, corresponding to $C_2 / C_5 \to 0$, is also shown for comparison. 
In all of our calculations, $C_2 = 1.0$.
One can observe the approach to the CSS result as $C_5$ increases.

Finally, we restore the explicit $\Xi\xleft(\Tsc{q}/Q,\eta\right)$ in the asymptotic term and 
calculate the $Y$-term according to Eq.~\eqref{eq:finalydef2} for two values of $Q$, one large and one small. 
The results are shown in Figs.~\eqref{fig:yterms}(a,b).
Here we use $C_5 = 1.0$ as a compromise between the various choices in Fig.~\ref{fig:asymplots} and to match with a 
common choice used in calculations like those of Ref.~\cite{Bozzi:2005wk}.   
For $Q = 20$~GeV (Fig.~\ref{fig:yterms}(a)), there is 
a region $1.0 \, {\rm GeV} \lesssim \Tsc{q} \lesssim 6.0 \, {\rm GeV}$ where the $Y$-term is a useful non-trivial
correction. Beyond about $\Tsc{q} \approx 6.0$~GeV, the $Y$-term simply approaches the $\fixo{}{}$ calculation (where the $W$-term vanishes).

Within our $W+Y$ method, the $Y$-term remains a reasonable correction for large $\Tsc{q}/Q$ even down to $Q = 2.0$~GeV, as shown in
Fig.~\eqref{fig:yterms}(b). There it forces a matching with the $\fixo{}{}$ calculation at $\Tsc{q} = O(Q)$, while it vanishes for small $\Tsc{q}$.

Note that, if the entire range of $\Tsc{q}$ up to order $Q$ is considered, then the treatment of the $Y$-term
plays an important role in describing the general shape of the $\Tsc{q}$-spectrum, particularly for the smaller $Q$ values. Indeed, for smaller $Q$, the $Y$-term 
appears to dominate the tail region. 
These observations highlight the importance of achieving well-constrained \emph{collinear} treatments of the large $\Tsc{q}$ region. 
Most likely, calculations of the fixed order term to rather high order should be included in implementations to adequately describe the large $\Tsc{q}$ behavior. For instance, Ref.~\cite{Daleo:2004pn} finds that 
order $\alpha_s^2$ fixed order calculations are needed to get
acceptable phenomenological success (see the comparison of curves in
Fig.~4 of Ref.~\cite{Daleo:2004pn}). Reference~\cite{deFlorian:2013taa} finds that threshold resummation corrections are also needed.

\section{Breakdown of Factorization in the Photoproduction Limit}
\label{sec:phot}
\begin{figure}
\centering
  \begin{tabular}{c@{\hspace*{10mm}}c}
    \includegraphics[scale=0.3]{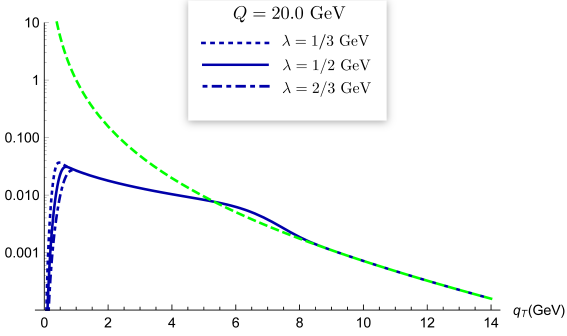} \\
    (a) \\
    \vspace{8mm} \\
    \includegraphics[scale=0.3]{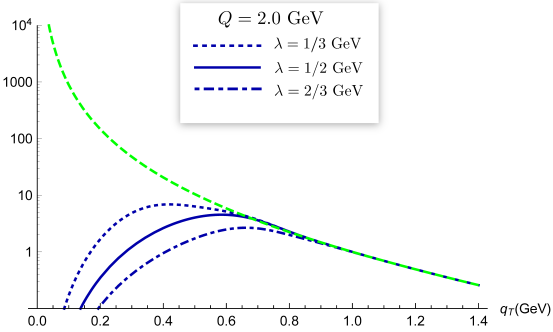}
    \\
    (b)
    \vspace{8mm} \\
    \includegraphics[scale=0.3]{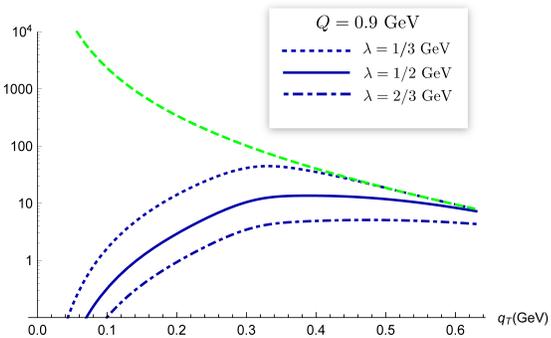}
    \\
    (c)
    \\[5mm]
   \end{tabular}
\caption{The $Y$-term calculated with $C_5=1.0$ and with the three values: $\lambda = 1/3$~GeV (blue dotted curves), $\lambda=1/2$~GeV (blue solid curves) 
and $\lambda=2/3$~GeV (blue dot-dashed curves). The green dashed curves show the fixed order calculations. Graph (a) is for $Q=20$~GeV, graph (b) is for $Q=2$~GeV, and 
graph (c) is for $Q=0.9$~GeV. (Color online. See text for further explanation.) In all cases, we take $x = 0.1$ and $z = 0.5$.
}
\label{fig:ytermslam}
\end{figure}
Of course, both TMD and collinear factorization theorems apply to the limit of a large hard scale $Q$;
part of the statement is that corrections to the factorized formulas are suppressed by powers of $m/Q$. Therefore, one expects factorization to 
work well in practice for very large $Q$ and to fail completely for $Q \to 0$, with the in-between region being less clear. In the SIDIS case, the $Q \to 0$ limit corresponds 
to photoproduction: $\gamma + P \to H + X$. If $Q$ is gradually decreased from some 
initially very large values, one expects uncertainties related to the general onset of non-perturbative physics beyond factorization 
to gradually increase.  

This is, of course, a standard and well-known aspect of QCD. 
The most obvious signal of the breakdown of perturbative QCD factorization is that $\alpha_s(Q)$ begins 
to blow up when $Q \to O(m)$. However, it is instructive to examine the transition from the solidly large $Q$ region to the $Q \to 0$ region in more detail. 
An analysis of the transition could guide applications of TMD factorization over a wide range of scales, aid in error analysis in applications, and provide intuition for how to match to truly non-perturbative physics.
For example, in the true photoproduction limit, it may be useful to switch to a physical picture more closely resembling a Regge exchange model~\cite{Kramer:1978xc}.

TMD factorization is most useful if there are distinct regions where 
i) $\Tsc{q} \lesssim O(m)$, where TMD correlation functions can be used,
and ii) $\Tsc{q} \sim O(Q)$, where collinear factorization applies. One way to test whether that is the case is to vary the $\lambda$ 
parameter of Eq.~\eqref{eq:yterm}. This controls the suppression of the $Y$-term for $\Tsc{q} < O(m)$, so varying it 
should have small or negligible effects on how one treats the perturbative $\Tsc{q}$-dependence at $\Tsc{q} \gg m$. 

Figure~\ref{fig:ytermslam}(a) confirms that this is true in our sample calculation for a very large value of $Q = 20$~GeV.  The graph shows the $Y$-term for several values of $\lambda$  
along with the $\fixo{}{}$ calculation (dashed) for comparison. The region where $1~{\rm GeV} \lesssim \Tsc{q} \lesssim 6~{\rm GeV}$ corresponds 
to a region where, roughly, $m \ll \Tsc{q} \ll Q$. Therefore, one could probably rely mostly on the $W$-term with its TMD pdfs and ffs to 
give a reasonable general description for $1~{\rm GeV} \lesssim \Tsc{q} \lesssim 6~{\rm GeV}$. 
However, $\Tsc{q}$ is still large enough in this region that $\Tsc{q}/Q$ power corrections to the $W$-term might not be totally 
negligible, so the $Y$-term is a useful and important correction to a $W$-term calculation in the moderate $\Tsc{q}$ region. Including it can 
enhance precision over a wide range of $\Tsc{q}$. In the $1~{\rm GeV} \lesssim \Tsc{q} \lesssim 6~{\rm GeV}$ region
the $Y$-term has negligible sensitivity to the exact value of $\lambda$ (so long as $\lambda = O(m)$). The $Y$-term is, therefore, unambiguous in 
its region of relevance, up to choices in $C_5$ and $\eta$. Moreover, variations in $C_5$ and $\eta$ can be understood in terms of higher order corrections. 
There is some residual sensitivity to $\lambda$ for $\Tsc{q} < 1.0$~GeV, but for $Q=20$~GeV,  $\Tsc{q} < 1.0$  definitely corresponds to 
a region where $\Tsc{q}/Q \ll 1$.  So, we are justified in simply ignoring the $Y$-term in the $\Tsc{q} \lesssim 1.0$ region.

In Fig.~\ref{fig:ytermslam}(b), we consider the lower $Q$ value of $2.0$~GeV,  where $Q$ is relatively small, but still 
large enough to hope that TMD factorization is still useful. The range of $\Tsc{q}$ as a fraction of $Q$ is the same as in Fig.~\ref{fig:ytermslam}(a). 
For this smaller $Q$ region, one might reasonably expect values of 
$\Tsc{q} \approx 0.2$~GeV to about $\Tsc{q} \approx 0.9$~GeV to qualify as $\Tsc{q} \ll Q$. However, $\Tsc{q}$ is still large 
enough here that concerns about $\Tsc{q}/Q$ power corrections from a $Y$-term are definitely warranted. As can be seen from 
the graph, the $Y$-term has significant uncertainties at intermediate $\Tsc{q}$ when $Q \sim 2.0$~GeV coming from the exact choice of $\lambda$. 
Nonetheless, the region of $\Tsc{q} \lesssim 0.2$~GeV corresponds to $\Tsc{q}/Q \lesssim 0.1$, so in the the smallest $\Tsc{q}$ region 
one may be confident in the applicability of factorization. Likewise, for $\Tsc{q} \gtrsim 0.9$~GeV, one may begin relying on 
the $\fixo{}{}$ calculation. Our $W+Y$ construction interpolates between these two descriptions. Therefore, it is reasonable to expect a fit to $Q=2.0$~GeV data 
to be qualitatively consistent with TMD factorization, even though 
uncertainties associated with $m/Q$-suppressed violations of factorization may begin to be more discernible. Said differently, 
at $Q \sim 2.0$~GeV, there may be a window of intermediate $m < \Tsc{q} < Q$ where $m/\Tsc{q}$ and $\Tsc{q}/Q$ are not both simultaneously small, yet we 
obtain a reasonable overall description by calculating the $\Tsc{q} \lesssim m$ behavior and the $\Tsc{q} \sim Q$ behavior and interpolating between the two. The only uncertainty then is 
in the exact nature or the interpolation. Notice, furthermore, that once a fit has been performed at $Q \sim 2.0$~GeV, any sensitivity to $\lambda$ 
automatically vanishes after evolution to large $Q$, as illustrated by Fig.~\ref{fig:ytermslam}(a). In other words, there is no disadvantage to 
optimizing fits at liberally low $Q$ since the limiting behavior at large $Q$ is unaffected.

To see the total breakdown of TMD factorization explicitly, we may push to even lower $Q$; in Fig.~\ref{fig:ytermslam}(c) we repeat 
the calculation of the $Y$-term for $Q = 0.9$~GeV, again over the same range of $\Tsc{q}$ as a fraction of $Q$. Here, $\Tsc{q}/Q$ corrections 
may be important already at $\Tsc{q} \sim 0.2$~GeV where transverse momentum dependence is still non-perturbative and the $Y$-term is totally 
controlled by the value of $\lambda$, and the functional form of $X(\Tsc{q})$.  A fit done in this region will likely be totally dominated by the (arbitrary, as far 
as TMD factorization is concerned) choice in $X(\Tsc{q})$ and the value of $\lambda$. Thus, as $Q$ drops below about $1$~GeV, 
TMD factorization begins to lose its predictive power and its usefulness. 

It is important to emphasize that the calculations in Figs.~\ref{fig:asymplots}--\ref{fig:ytermslam} are meant to demonstrate 
general features only. To gain a true understanding of the moderate $Q$ region and the transition to the $Q \to 0$ limit, 
up-to-date collinear pdfs and ffs are needed for all partonic channels, higher order perturbative calculations including 
flavor thresholds should be included, and a $W$-term with a specific parametrization for non-perturbative $\Tsc{q}$-dependence is 
needed. 

The region of $Q$ of order a few GeV can likely be enhanced by extensions of factorization to higher twist~\cite{Arleo:2010yg} and/or by using new 
non-perturbative correlation functions that treat kinematics exactly like fully unintegrated pdfs~\cite{Collins:2005uv,Collins:2007ph,Rogers:2008jk}.

\section{Summary}
\label{sec:con}

We conclude by summarizing the logic of our modified  
$W + Y$ construction. 

TMD factorization applies for small $\Tsc{q}\ll Q$ and degrades in
accuracy as $\Tsc{q}$ increases.  In contrast, collinear factorization
applies when $\Tsc{q} \sim Q$ and also to the cross section integrated
over all $\T{q}$; its accuracy on the differential cross section
degrades as $\Tsc{q}$ decreases.  The standard $W + Y$ prescription
was arranged to apply also for intermediate $\Tsc{q}$; in particular
it keeps full accuracy when $m \ll \Tsc{q} \ll Q$, a situation in
which both pure TMD and pure collinear factorization have degraded
accuracy\footnote{What we have
    in mind here is that the errors in TMD factorization include a
    power of $\Tsc{q}/Q$ as well as a power of $m/Q$, while the error
    in collinear factorization is a power of $m/\Tsc{q}$.  Now an
    optimal blend of TMD and collinear factorization can have an error
    of a particular power of $m/Q$ uniformly in $\Tsc{q}$.  But,
    because of the $\Tsc{q}/Q$ and $m/\Tsc{q}$ errors in each
    individual kind of factorization, either one of these by itself
    has much worse accuracy than the blend, when $m \ll \Tsc{q} \ll
    Q$.}. However, it did not specifically address the issue of
matching to collinear factorization for the cross section integrated
over $\T{q}$. 

Furthermore, for the $\Tsc{q} \gtrsim Q$ and $\Tsc{q} \lesssim m$
regions, the CSS $W+Y$ formalism as it stands does not robustly revert
to the $\fixo{}{}$ or $\TT{}{}$ terms alone.  A variety of methods for
dealing with these and related issues exists in the literature (see
Sec.~\ref{sec:principles}), but they usually appear at the level of
implementations rather than in the formalization itself.  We have
synthesized components from these previous approaches into a
relatively compact prescription.

With our method, the redefined $W$ term allowed us, in
Sec.~\ref{sec:together},
to construct a relationship between between integrated-TMD-factorization formulas and standard collinear factorization formulas, with 
errors relating the two being suppressed by powers of $\bmin/\bmax \sim 1 /Q$. Importantly, the exact definitions of the TMD pdfs and ffs are unmodified from the usual ones of factorization derivations 
(e.g., Eqs.~(13.42,13.106) of Ref.~\cite{Collins:2011qcdbook}).
  We preserve transverse-coordinate space version of the $W$ term,
  but only modify the way in which it is used.  Thus the derivation
  of TMD factorization is preserved, and we have only changed the way
  in which the ingredients are assembled into a formula for the cross
  section.
Finally, the standard CSS formalism, with its 
more standard $W +Y$ construction, is automatically 
recovered in the limit of very large $Q$. Having organized a systematic treatment of the matching between $W$ and $Y$ terms, we may begin 
to incorporate physically motivated considerations (e.g., similar to the momentum rescaling of Ref.~\cite{Guzzi:2013aja}) into the construction of 
specific functional forms for $\Xi$ and the choice of $C_5$.

This paper has dealt only with unpolarized cross sections. However, we expect analogous reasoning to apply 
when polarization is taken into account. In such cases, the connection between large and small $\Tsc{q}$ 
is more subtle because power counting can differ at large and small $\Tsc{q}$ depending on the specific polarization 
observable under consideration. This is discussed extensively in Ref.~\cite{Bacchetta:2008xw}. To implement steps analogous to those we have presented here, one most likely needs to 
consider $\Tsc{q}$-weighted integrals of cross sections or weighting by Bessel functions~\cite{Boer:2011xd}. Such studies may 
prove to be especially interesting in how they relate correlation functions of different twist. 

Many planned applications of TMD factorization depend crucially on the ability to control matching between 
perturbatively large and non-perturbatively small $\Tsc{q}/Q$. This is especially the case for phenomenological studies where the 
shape of the distribution and the possible presence of a non-perturbative $\Tsc{q}$-tail is a central question, such as in 
studies of flavor-dependence in TMDs~\cite{Signori:2013mda}, or a potential difference between sea 
and valence quark intrinsic transverse distributions~\cite{Schweitzer:2012hh}.  (See also Fig.~1 of Ref.~\cite{Aidala:2014hva}.)
With the method of this paper, it is possible in principle to interface 
the full $W+Y$ TMD construction with generalized parton model approaches to phenomenology like Refs.~\cite{Signori:2013mda,Anselmino:2013lza,Melis:2014pna} -- a step that 
we leave for future work.

We plan to next apply our enhanced $W + Y$ construction in phenomenological studies. 
In particular, given the rather low $Q$-values typical of SIDIS experiments, we expect analyses of unpolarized SIDIS 
to benefit from the greater control over the transition from small $\Tsc{q}$ to $\Tsc{q} \sim Q$. 


\begin{acknowledgments}
D.~B.~Clark provided numerical 
help on calculations performed in an earlier version of this paper. We thank D.~Boer and M.~Diehl for many useful 
comments and discussions regarding the text. We also thank C.~Aidala, C.~Courtoy, O.~Garcia and P.~Nadolsky for general conversations regarding factorization.  
  This work was supported by DOE contracts No.\ DE-AC05-06OR23177
  (A.P., T.R., N.S., B.W.), 
  under which Jefferson Science Associates, LLC operates Jefferson
  Lab,
  No.\ DE-FG02-07ER41460 (L.G.), and No.\ DE-SC0008745 (J.C.), 
  and by the National Science Foundation under Contract No. PHY-1623454 (A.P.).
\end{acknowledgments}

\appendix

\section{Proof that the standard CSS $\tilde{W}(\Tsc{b})$ vanishes for $\Tsc{b} \to 0$}
\label{sec:Wzero}

We obtain the small $\Tsc{b}$ behavior of $\tilde{W}(\Tsc{b})$
from Eq.\ (\ref{eq:finalevolved}), showing that it behaves
approximately as a calculated power of $\Tsc{b}$ when $\Tsc{b} \to 0$.
In this limit $\mubstar \to \infty$, and so because of asymptotic
freedom, $\alpha_s(\mubstar) \to 0$.  The dominant behavior is then
given by the effect of the lowest order term in $\gamma_K$.  

To see this explicitly, we compute the derivative of
$\tilde{W}(\Tsc{b})$ with respect to $\ln \Tsc{b}$:
\begin{align}
  \frac{ \partial \tilde{W}(\Tsc{b}) }
       { \partial \ln \Tsc{b} }
  ={}& -4C_F \frac{\alpha_s(\mubstar)}{\pi}
       \ln\xleft( \frac{Q}{\mubstar} \right)
       \tilde{W}(\Tsc{b})
\nonumber\\&
   + H \times (\mbox{ff and pdf factor}) \times e^{-S}
\nonumber\\&~~
    \times 
    \left[
      O(\alpha_s(\mubstar))
      +
      O(\alpha_s(\mubstar)^2) \ln\xleft( \frac{Q}{\mubstar} \right)
    \right],
\label{eq:W.diff.b}
\end{align}
where we used $\gamma_K = 2C_F \alpha_s/\pi + O(\alpha_s^2)$.  The
factor $e^{-S}$ represents the exponential factors on the last two
lines of (\ref{eq:finalevolved}).  In (\ref{eq:W.diff.b}), the
subleading terms on the last line come from a combination of DGLAP
evolution of the collinear pdf and ff, from differentiating the
$\tilde{C}$ factors, and from differentiating the terms in $S$ other
than the leading order term in $\gamma_K$.

Now, when $\mu\to\infty$,
\begin{equation}
   \frac{\alpha_s(\mu)}{\pi} 
   \sim \frac{ 2 }{ \beta_0 \ln(\mu/\Lambda_{\rm QCD})},
\end{equation}
where $\beta_0= 11 -2n_f/3$.
Therefore we get
\begin{align}
  \frac{ \partial \tilde{W}(\Tsc{b}) }
       { \partial \ln \Tsc{b} }
  ={}& \frac{ 8C_F }{ \beta_0 } ~ \tilde{W}(\Tsc{b})
\nonumber\\&
       + O\xleft( \frac{ H \times (\mbox{ff and pdf factor}) e^{-S} }
                       {\ln(\Tsc{b}\Lambda)}
          \right),
\end{align}
as $\Tsc{b} \to 0$.  Hence, $\tilde{W}$ itself has approximately a
power behavior:
\begin{align}
   \tilde{W}(\Tsc{b} \to 0,Q) = \Tsc{b}^{a} 
   \times (\mbox{logarithmic corrections}) ,
\label{eq:bdep2}
\end{align}
where $a = 8 C_F / \beta_0$. 

Note that the pdf and ff factors in (\ref{eq:finalevolved})
themselves go to zero as $\mubstar \to \infty$, but more slowly than a
power.

\section{Low Order Asymptotic Term Expressions}
\label{sec:asy}
Calculations for the asymptotic term in a usual formalism can be found in, for example, Refs.~\cite{Nadolsky:1999kb,Koike:2006fn}. Using a non-zero 
$C_2/C_5$ we obtain from Eqs.~\eqref{eq:logrep1}--\eqref{eq:logrep2}:
\begin{widetext}
\begin{align}
\Xi\xleft(\frac{\Tsc{q}}{Q},\eta\right) \frac{ \alpha_s(\muQ)}{2\pi sx_A} \frac{C_2 b_0}{\Tsc{q} \muQ C_5} & \sigma_0  \sum_{q,\bar q} e_q^2\Big[ 2 f_q(x_A,\muQ) D_q(z_B,\muQ) \Big(C_F \left[ K_1 \xleft( \frac{C_2 \Tsc{q} b_0}{C_5 \mu_Q} \right) \ln \xleft( \frac{C_2 \muQ}{C_5 \Tsc{q}} \right) + \right. \nonumber \\ 
                             &   \left.  
 + K_1^{(1)}\xleft( \frac{C_2 \Tsc{q} b_0}{C_5 \muQ} \right) \right]-\left(\frac{3}{2} + \ln(C_2^2) \right) C_F K_1 \xleft( \frac{C_2 \Tsc{q} b_0}{C_5 \muQ}\right)\Big) \nonumber \\
 & + K_1 \xleft( \frac{C_2 \Tsc{q} b_0}{C_5 \muQ} \right) \left( f_q(x_A,\muQ)\otimes P_{qq}^{(0)} + f_g(x_A,\muQ)\otimes P_{qg}^{(0)}\right) D_q(z_B,\muQ) \nonumber \\
 & + K_1 \xleft( \frac{C_2 \Tsc{q} b_0}{C_5 \muQ} \right) f_q(x_A,\muQ) \left( D_q(z_B,\muQ)\otimes P_{qq}^{(0)} + D_g(z_B,\muQ)\otimes P_{gq}^{(0)}\right)  \Big] \, .
 \end{align}
 \end{widetext}
where
\begin{align}
\sigma_0= \frac{4\pi\alpha^2_{\rm em}}{s x_A y^2} \left(1-y+\frac{y^2}{2}\right) \, , 
\end{align}
and leading order splitting functions are
\begin{align}
P_{qq}^{(0)} &= C_F \left[ \frac{1+x^2}{(1-x)_+} +\frac{3}{2} \delta(1-x) \right]  \nonumber \\
P_{gq}^{(0)} &= C_F  \frac{1+(1-x)^2}{x} \nonumber \\
P_{qg}^{(0)} &= T_R  (x^2+(1-x)^2) 
\end{align}
Note that $\asnew{}{}$ is differential in $d x_A  d z_B d Q^2   d{q}_{T}^2$. The convolution $\otimes$ is defined in a standard way:
\begin{align}
f_q(x_A,\mu) \otimes P_{qq}^{(0)} &\equiv \int_{x_A}^1 \frac{dx}{x} \ f_q(x,\mu) \ P_{qq}^{(0)}\xleft(\frac{x_A}{x}\right)
\end{align}

\bibliography{tcr}

\end{document}